\documentclass[twocolumn, twocolappendix]{aastex701}

\usepackage{amsmath}
\usepackage{bm}

\newcommand{\LIRA}{LIRA, Observatoire de Paris, Universit\'e PSL, Sorbonne Universit\'e, Universit\'e Paris Cit\'e, CY Cergy Paris Universit\'e, CNRS, 92190 Meudon, France}

\begin{document}

\title{Tearing Driven Reconnection: Energy Conversion Involving Firehose Kinetic Instabilities (2D Hybrid M\"obius Simulations)}

\author{Etienne Berriot}
\affiliation{\LIRA}
\email[show]{etienne.berriot@obspm.fr}

\author{Petr Hellinger}
\affiliation{Astronomical Institute of the Czech Academy of Sciences, Prague, Czechia}
\affiliation{Institute of Atmospheric Physics of the Czech Academy of Sciences, Prague, Czechia}
\email{petr.hellinger@asu.cas.cz}

\author{Olga Alexandrova}
\affiliation{\LIRA}
\email{olga.alexandrova@obspm.fr}

\author{Alexandra Alexandrova}
\affiliation{Institute for Humanity's Unified Development, Geneva, Switzerland}
\email{sasha.alexandrova@gmail.com}

\author{Pascal D\'emoulin}
\affiliation{\LIRA}
\email{pascal.demoulin@obspm.fr}

\begin{abstract}
This study focuses on energy conversion related to tearing-driven magnetic reconnection in the context of weakly collisional astrophysical plasmas.
We present results from a two-dimensional hybrid particle-in-cell simulation employing novel periodic conditions with a topology akin to the M\"obius strip, which double the computation efficiency as compared to regular periodic conditions.
Evaluation of the ion electric work rate ($\bm{j}_i \cdot \bm{E}$) and pressure strain interaction ($\mathbf{P}_i : \bm{\nabla u}_i)$ shows that most of the energy conversion occurs during the nonlinear phase of the instability, where magnetic energy is transferred towards ion kinetic energy (bulk outflows) and internal energy (heating).
These energy conversion rates are of the same order but inhomogeneous.
Heating predominantly occurs within the magnetic islands, while near the X-points, nearly the same amount of magnetic energy is transferred to bulk plasma flow and heating.
The reconnected plasma moreover exhibits an ion temperature higher parallel than perpendicular to the local magnetic field $\bm{B}$.
This temperature anisotropy is sustained by the islands' contraction, but eventually gets regulated by kinetic firehose instabilities (parallel and oblique), and/or firehose-like processes, whose main effect is to redistribute the internal energy from the parallel to the perpendicular direction.
\end{abstract}


\section{Introduction} \label{sec:intro}

Astrophysical plasmas are prone to the development of magnetic structures, such as current sheets: thin layers where an important magnetic shear gives rise to intense electric currents.
Current sheets are ubiquitous in weakly collisional astrophysical plasmas like the solar wind, and are the places where magnetic reconnection can occur.
Magnetic reconnection is a fundamental process in plasma physics, which leads to sudden changes in the magnetic field's topology at so called X-points, and causes important transfer of magnetic energy towards bulk flow, heating, and particle acceleration.
Elongated current sheets may moreover be unstable to the tearing (or plasmoid) instability \citep[see, e.g.][]{Furth_1963_resitive_instabilities, Loureiro_2007_plasmoids_chains, Ji_2011_phase_diagram_reconnection, Pucci-Velli_2014_ideal_tearing}.
The tearing instability spontaneously triggers several magnetic reconnection events at different sites (X-points) along a current sheet, and generates closed magnetic structures: magnetic islands (also called plasmoids) in between, as evidenced by both simulations \citep[e.g.][]{Karimabadi_1999_plasmoid} and observations \citep{Alexandrova_2016_interacting_X-lines}.
These magnetic islands are highly dynamic and may undergo coalescence, driving other magnetic reconnection events between them as they merge, thus bringing even more complexity to this plasma process \citep[e.g.][]{Pritchett_1992_coalescence, Markidis_2012_plasmoid_chain_coalescence, Nakamura_2023_coalescence}.
Magnetic reconnection and the tearing instability are moreover involved in many astrophysical phenomena, such as solar flares and coronal mass ejections \citep{Parker_1963_flare_reco, Liu_2013_plasmoid_solar_flares, Priest_2014_book_MHD_Sun, Shibata_2016_fractal_reco_review, Arnold_2021_elec_acceleration_solar_flares_tearing_simu}, dynamics of the Earth magnetosphere \citep{Dungey_1961_Dungey_cycle,Fuselier_2024_review_earth_global_reconnection}, or production of high energy electromagnetic emissions in plasmas related to compact objects like pulsar winds \citep[e.g.][]{Cerutti_2013_tearing_2D,Philippov_2019_pulsar_wind_tearing}, and play an important role in turbulence \citep[e.g.][]{Servidio_2009_reconnection_turbulence, Franci_2017_reconnection_sub-ion_turbulence, Starwarz_2024_review_interplay_turbulence-reconnection}.
Numerous simulation studies have also shown the development of turbulence driven by magnetic reconnection \citep{Daughton_2011_3D_reconnection_turbulence_electrons, Daughton_2014_3D_reconnection_turbulence_rec-rate, Pucci_2018_turbulence_plasmoid, Lapenta_2020_local_turbulence_3D_reconnection, Adhikari_2020_2D_reconnection_turbulence_spectra, Adhikari_2024_scale-filtering_turbulence_plasmoid_2D, Starwarz_2024_review_interplay_turbulence-reconnection}


Signatures of magnetic reconnection can be studied with the help of in-situ measurements, such as in the solar wind and in the Earth magnetosphere \citep{Gosling_2012_review_reco_SW, Paschmann_2013_review_in-situ_reco}.
In particular, the MMS mission \citep{Burch_2016_MMS_mission} has been able to provide valuable information on the particles' dynamics by directly probing the reconnection's ion and electron diffusion regions \citep[see e.g.][and references therein]{Norgren_2025_review_reconnection_particle_dynamics}.
Moreover, the new missions Parker Solar Probe \citep{Fox_2016_Parker_Solar_Probe_science_goals} and Solar Orbiter \citep{Muller_2020_Solar_Orbiter_science_goals} are going closer than ever to the Sun, allowing to study the dynamics of the young solar wind, where magnetic reconnection exhausts are abundantly observed \citep{Phan_2020_PSP_E1_reco, Phan_2021_HCS_reco_PSP, Phan_2024_PSP_HCS_coalescence, Eriksson_2024_PSP_reco, Lewis_2026_HCS_reconnection_FR}.
However, in-situ measurements only provide physical parameters as 1 dimensional cuts (temporal series) inside a 3-dimensional and temporally evolving plasma, making it difficult to disentangle contributions from different phenomena.
This constraint can be somewhat alleviated by the use of several spacecraft, such as in the case of MMS.
These indeed allow estimation of spatial gradients, which can be used to evaluate quantities associated with energy conversion and transfer between scales in the plasma \citep{Manzini_2024_cascade-dissipation_MMS, Dahani_2024_energy_conversion_reconnection_MMS}.
On the other hand, numerical simulations have the advantage of providing the system's evolution over the entire domain, making it possible to do global studies of the plasma dynamics.

Regarding numerical simulations, the tearing instability in weakly collisional plasmas can be studied using a fluid approach, such as resistive, or Hall-MHD  \citep[e.g.][]{Bhattarcharjee_2009_tearing_simu_resistive_MHD, Papini_2019_sec_tearing_Hall-MHD, Huang-Bhattacharjee_2024_3D_tearing_Hall}.
These however do not include potentially important kinetic effects \citep[e.g.][]{Daughton_2009_tearing_transition_collisional-kinetic, Aunai_2011_simu_hybrid_reconnection_kinetics, Cozzani_2021, Norgren_2025_review_reconnection_particle_dynamics, Graham_2025_waves_reconnection}.
The fine dynamics, as well as the energy conversion processes, related to the tearing instability and magnetic reconnection depend on the plasma regime and is not yet fully understood, although progress is being made, in particular with the help of recent advances in the domain of numerical simulations \citep[see][and references therein]{Ji_2022_exascale_reconnection}.

An important parameter characterizing the magnetic reconnection efficiency is the reconnection rate, traditionally defined in a 2D geometry as the temporal change of magnetic flux at a reconnection point.
This rate, or proxies of it, can therefore be evaluated using the out-of-plane vector potential in simulations \citep[e.g.][]{Papini_2019_Hall-MHD_turbulence}, or through the out-of-plane electric field near the X-point $E^X_z$ for both simulations and observations.
Remarkably, when kinetic effects dominate over resistivity, the transfer of magnetic flux is significantly faster than in the purely resistive case \citep{Birn_2001_GEM}, and there seem to be a universal value of the reconnection rate proxy $E^X_z / (v_{A,0} B_0) \sim 0.1$, with $v_{A,0}$ and $B_0$ the upstream Alfv\'en speed and magnetic field amplitude \citep[see][and references therein]{Cassak_2017_review_01_rec_rate}.

In weakly collisional plasmas, the conversion rates between the different energy types (kinetic, internal and electromagnetic) can be quantified by the electrical work rate and the pressure-strain interaction.
The electrical work rate quantifies the exchanges between the electromagnetic field and the particles, while the pressure-strain interaction couples the particles kinetic and internal energies, and may work as an effective dissipation channels \citep{Del-Sarto_2016_anisotropy_velocity_shear, Yang_2017_pressure-strain, Yang_2022_PS_dissipation, Hellinger_2022_ion-scale_transition_turbulence_PS, Arro_2022_energy_transfer_turbulence, Cassak-Barbhuiya_2022_PS_interaction_I, Yang_2024_electron_dissip_EM_work}.

Astrophysical plasmas, such as the solar wind, are typically non-collisional magnetized medium, and therefore tend to develop temperature anisotropies.
These have been shown to affect the growth rate of the tearing instability and associated reconnection rates \citep{Matteini_2013_tearing_anisotropic, Gingell_2015_tearing_3D_anisotropy}.
The temperature anisotropies are however regulated (constrained) by instabilities, such as the proton parallel and oblique kinetic firehose instabilities when $T_{i \parallel} > T_{i \perp}$ \citep{Parker_1958_FH_instability, Gary_1976_linear_FH, Gary_1998_parallel_FH, Hellinger-Matsumoto_2000_oblique_FH}, and by the mirror and proton cyclotron instabilities when $T_{i \perp} > T_{i \parallel}$ \citep{Gary_1992_mirror_ion-cyclotron, Gary_1994_proton_cyclotron_anisotropy-beta, Southwood-Kivelson_1993_mirror_linear, Pantellini_1995_NL_mirror}, with $T_{i \parallel}$ and $T_{i \perp}$ the proton temperature parallel and perpendicular to the magnetic field $\bm{B}$, respectively.
When excited, these instabilities redistribute the plasma internal energy and re-isotropize the temperature.

Magnetic reconnection outflows generally exhibit parallel temperature anisotropies.
This is linked to a Fermi-like process that can accelerate both electrons \citep{Drake_2006_electron_acc_contracting_island, Egedal_2013_electron_anisotropy_trapping_reconnection} and protons \citep{Drake_2010_proton_Fermi_acc_firehose} in the outflows and contracting magnetic islands during the non-linear stage of the tearing instability.
These temperature anisotropies not only have an influence on the X-line and exhausts properties, but can moreover become important enough to trigger the firehose instabilities \citep[e.g.][]{Karimabadi_1999_plasmoid, Drake_2010_proton_Fermi_acc_firehose, Liu_2011_reconnection_anisotropy_Riemann, Le_2014_PIC_reconnection_firehose, Ohia_2015_reconnection_bi-fluid_electron_anisotropy-firehose, Burgess_2016_reconnection_anisotropy_ion_firehose}.

Signatures of the kinetic parallel proton firehose instability at the edges of a magnetic island, have been studied by \citet{Alexandrova_2020_reconnection_firehose} with the help of both in-situ observations and PIC simulation.
Kinetic simulations of the tearing instability seem to indicate that the total parallel temperature anisotropy $T_{\parallel} > T_{\perp}$ also regulates the magnetic islands contraction due to the reduced effective magnetic tension, with more effect in higher plasma $\beta$ regimes due to the marginal firehose stability condition being reached for lower anisotropy values \citep{Schoeffler_2011_firehose_tearing_islands_growth}.

The goal of this study is to investigate energy conversion, as well as the effects of the temperature anisotropy and related kinetic instabilities (firehose), during non-linear stages of the tearing process.
We study results from a two-dimensional hybrid particle-in-cell (PIC) simulation, with a novel approach of modified periodic M\"{o}bius boundary conditions, of a current sheet with no guide field.
In Section \ref{sec:simu_setup}, we describe the numerical setup and associated boundary conditions.
We then give an overview of the global system evolution during its linear and non-linear stages in Section \ref{sec:tearing_global_evo}.
In Section \ref{sec:nrj_conv}, we focus on the energy evolution and conversion within the plasma.
Then, in Section \ref{sec:firehose_island}, we show evidence for the presence of ion firehose instabilities, or similar processes in an inhomogeneous plasma, during the non-linear stages of the tearing instability and islands formation/contraction.
Finally, we discuss our results and conclude in Section \ref{sec:conclusion}.

\section{Simulation Setup and M\"obius Boundary Conditions} \label{sec:simu_setup}


We study a two-dimensional hybrid PIC plasma simulation, where ions (a single-specie here) are treated as macro-particles and electrons as a massless isothermal charge neutralizing fluid.
The simulation has been done using the code CAMELIA\footnote[1]{See also \hyperlink{https://space.asu.cas.cz/~helinger/camelia.html}{https://space.asu.cas.cz/\string~helinger/camelia.html}.}
\citep{Franci_2018_CAMELIA}, based on the Code Current Advance Method and Cyclic Leapfrog \citep[CAM-CL,][]{Matthews_1994_CAM-CL}.
The domain is discretized with $4096 \times 2048$ ($x \times y$) cells, each having a size of $1/8 \; d_i$, with $d_i$ the background ion inertial length.
There is a small explicit resistivity  $\eta = 4 \times 10^{-4} \; \mu_0 \, d_i^2 \, \Omega_{ci}$, and 2048 macroparticles per cell.
The background ion and electron plasma betas have been set as $\beta_{0, i} = \beta_{0, e} = 0.25$.

\begin{figure}[t!] 
    \resizebox{\hsize}{!}{\includegraphics{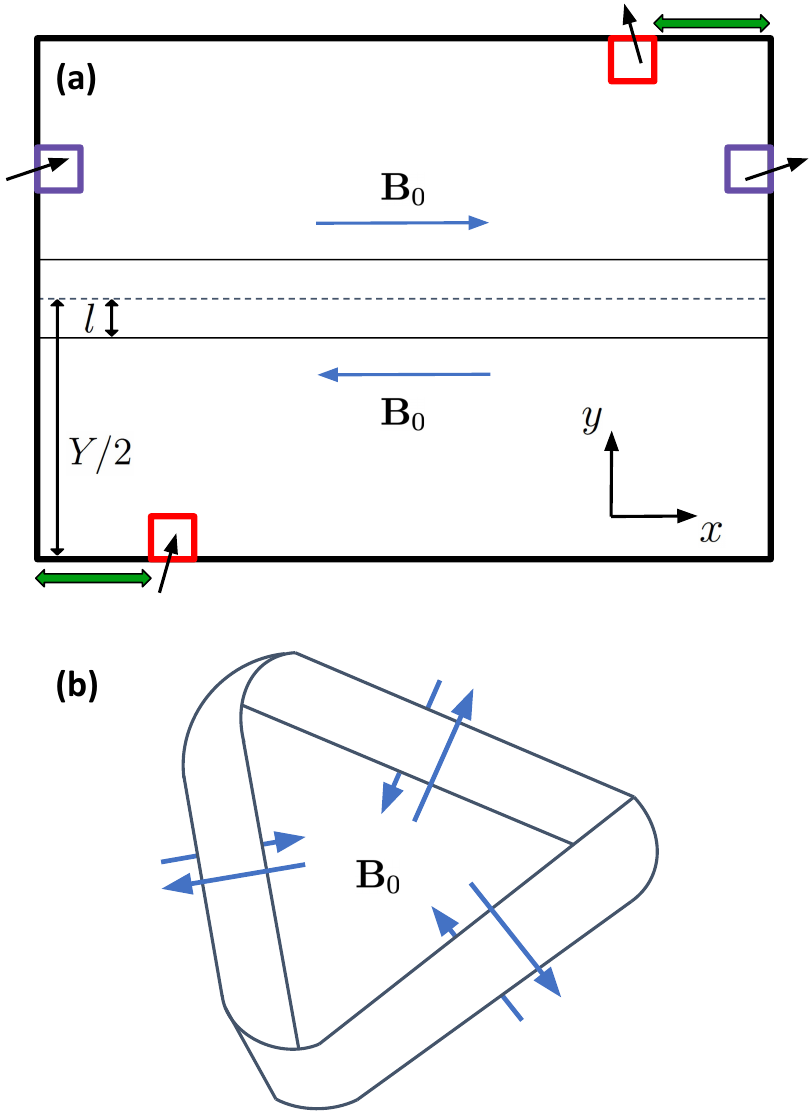}}
    \caption{\label{fig:simu_domain_schematic}
    Schematic views of the simulation domain and associated M\"obius boundary conditions.
    (a): 2D simulation domain, with the initial magnetic field $\bm{B}_0$ depicted by blue arrows.
    The boundaries are periodic in $x$, and have M\"obius periodic conditions in $y$.
    The positions of a particle (with velocity indicated by the black arrow) exiting and re-entering the domain are depicted by purple cells for the $x$ boundary and red cells for the $y$ boundary.
    (b): Reversal of the magnetic field vector along a M\"obius strip.
    }
    \end{figure}  

Figure \ref{fig:simu_domain_schematic}(a) shows a schematic of the 2D simulation domain, of total lengths $X$ and $Y$ in the $x$ and $y$ directions, respectively.
The initial magnetic field $\bm{B}_0$ (blue arrows) presents a current sheet of total thickness $2 \, l$ in the center of the domain.
The boundaries are periodic in $x$, as depicted by the purple cell representing the position of a particle (with velocity indicated by the black arrow) leaving and re-entering the domain through the vertical boundaries.
The up ($y = Y/2$) and down ($y = -Y/2$) boundaries have M\"obius periodic conditions, and the position of a particle leaving and re-entering the domain through the vertical boundary is indicated by the red cells.

The  M\"obius periodic boundary conditions are defined as follows.
For a macroparticle with coordinates ($x_p$, $y_p$) leaving the simulation domain with $y_p>Y/2$ or $y_p<-Y/2$, the new coordinates are assigned as:
\begin{eqnarray}
y_p > Y/2 \Rightarrow \left\{
    \begin{array}{ll}
        y_p \rightarrow y_p - Y \\
        x_p \rightarrow X - x_p
    \end{array}
\right.
\\
y_p < -Y/2 \Rightarrow \left\{
    \begin{array}{ll}
        y_p \rightarrow y_p + Y \\
        x_p \rightarrow X - x_p,
    \end{array}
\right.
\end{eqnarray}
as represented in Figure \ref{fig:simu_domain_schematic}.
Moreover, for any vector $\bm{V}$, the $x$ and $z$ components, $V_x$ and $V_z$, also obey the M\"obius boundary condition:
\begin{eqnarray}
    V_x (x, y=Y/2) &= - V_x (X - x, y=-Y/2) \nonumber \\
    V_z (x, y=Y/2) &= - V_z (X - x, y=-Y/2). \label{eq:Mobius_vector}
\end{eqnarray}
Particles leaving through the M\"obius boundary also have their velocity changed according to Equation (\ref{eq:Mobius_vector}).
Figure \ref{fig:simu_domain_schematic}(b) depicts how a vector, such as the magnetic field, reverses halfway along a M\"obius strip.

The M\"obius boundary conditions allow to only have a single current sheet in the domain instead of the two required when using standard periodic boundary conditions, essentially dividing the computational cost by two.
Over the studied simulation, the reconnecting current sheet and growing structures remain far enough from the $y = \pm Y/2$ boundaries that there is no other noticeable effect during the system evolution.
Otherwise, for a smaller domain's vertical extension, the current sheet would interact with itself, changing the overall system's dynamics, as would do two current sheets interacting with each other in a domain with regular periodic boundary conditions.
We note that analogous boundary conditions were independently developed and tested by \citet{Xia-Swisdak_2024_Klein_bottle_boundaries}, obtaining similar results to those of simulations with double periodic boundary conditions.

The simulation is initialized as a single current sheet in the center of the domain ($y=0$) in a Harris pressure equilibrium \citep{Harris_1962}, with no guide field and a homogeneous isotropic temperature with Maxwellian velocity distribution functions for the ions.
The ions are protons only, so quasi-neutrality gives $n_i = n_e = n$, with $n_i$ and $n_e$ the ion and electron density, respectively.
The magnetic field and density profiles are therefore such that
\begin{align}
    B_x (t=0, y) &= B_0 \tanh (y/l) \\
    n(t=0, y) &= n_0 \left( 1 + \frac{1}{\beta_{0, i} + \beta_{0, e}} \frac{1}{\cosh^2(y/l)} \right),
\end{align}
with the following quantities outside the current sheet: magnetic field $B_0$, plasma density $n_0 \neq 0$, ion and electron plasma background beta $\beta_{0, (i,e)} = n_0 k_B T_{0, (i,e)} / (B_0^2 /2 \mu_0)$, where $T_{0, i}$ and $T_{0, e}$ are the initial ion and electron temperatures.
We remind that $T_{0, i} = T_{0, e}$ and that the electrons are isothermal so $T_e = T_{0, e} \, \forall (t, x, y)$.
The current sheet half thickness is set as $l = 0.5 \, d_i$. 
We do not include any explicit initial background perturbation, so the tearing instability starts growing from the simulation numerical noise.

\section{Tearing Instability Onset and Evolution} \label{sec:tearing_global_evo}

\subsection{From Linear to Non-Linear Stage}

\begin{figure}[t!] 
    \resizebox{\hsize}{!}{\includegraphics{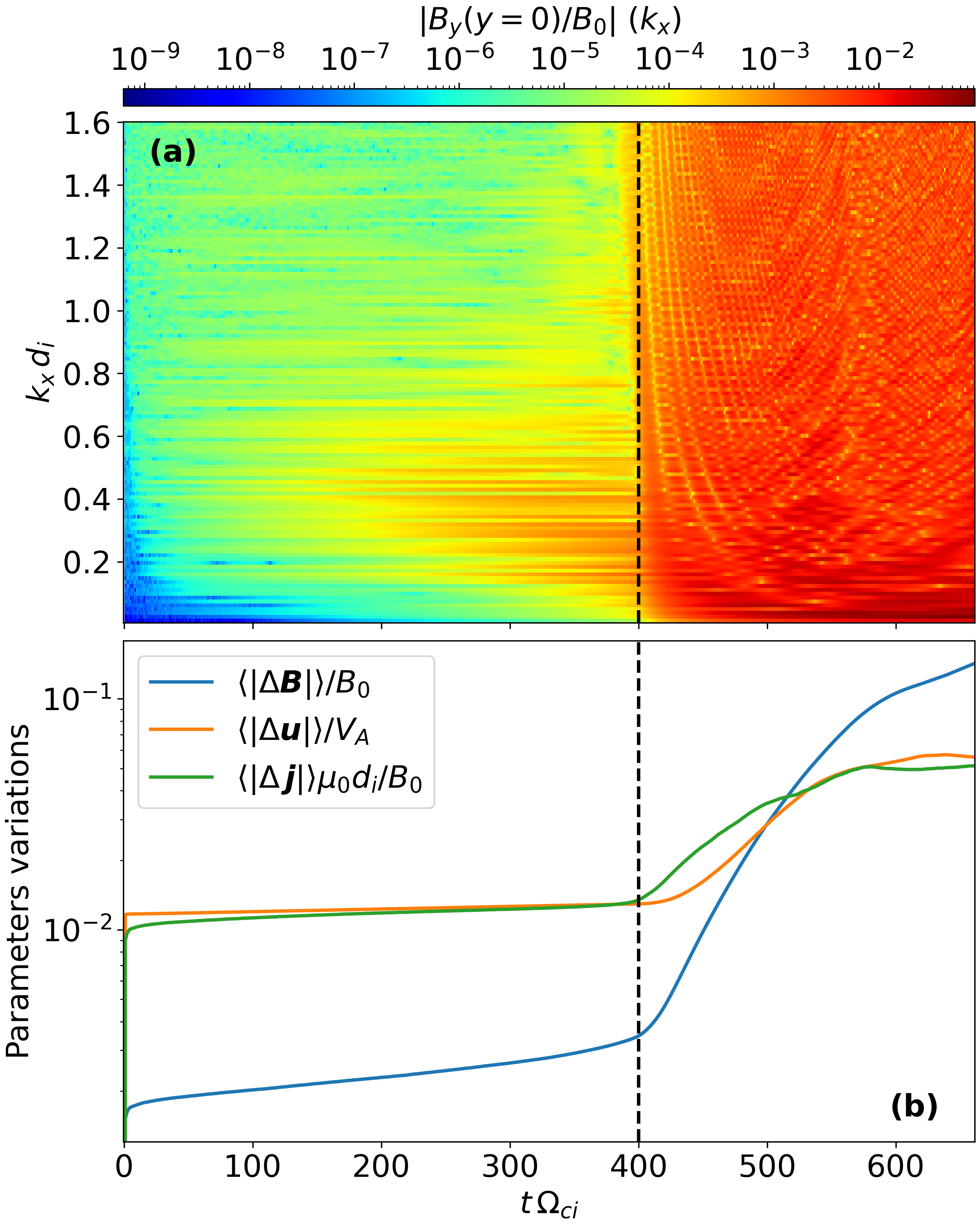}}
    \caption{\label{fig:FFT_By_t}
    Temporal evolution of the magnetic field fluctuations within the current sheet, and plasma variations averaged over the domain.
    Panel (a) shows the evolution of the magnetic field component $B_y$ in the center of the current sheet, as function of the wave number $k_x \, d_i < 1.6$.
    Colors indicate the amplitude of $|B_y (y=0)/B_0| (k_x)$ decomposed over $k_x$ using a fast Fourier transform (FFT) in the $x$ direction along the current sheet center.
    Panel (b) shows the, spatially averaged, variations of the magnetic field $\bm{B}$, plasma bulk velocity $\bm{u}$, and current density $\bm{j}$.
    The vertical dashed black line ($t = 400 \; \Omega_{ci}^{-1}$) indicates the transition from the linear to non-linear tearing instability.
    }
\end{figure}  

We first focus on the evolution of the magnetic field fluctuations within the current sheet during the tearing instability.
Figure \ref{fig:FFT_By_t}(a) displays the temporal evolution of the $B_y$ component in Fourier space $|B_y(y=0)/B_0| (k_x)$, the magnetic field reconnected component perturbations amplitude in the center of the current sheet, for $k_x \, d_i < 1.6$.
For $t < 400 \, \Omega_{ci}^{-1}$ there is a nearly linear stage of the instability, with perturbations at different wavenumbers growing mostly independently from one another.
As the instability evolves, low wavenumbers $k_x \, d_i \lesssim 0.6$ develop the most important perturbations, which then get to similar amplitudes by the end of the linear stage around $t = 400 \, \Omega_{ci}^{-1}$.
Inspection of $B_y$ in the real space (not shown) indeed revealed several X-points along the current sheet, with different reconnected field amplitudes and no clear dominant periodicity, as expected from the lack of a singular dominant wavenumber during this stage.
At times $t > 400 \, \Omega_{ci}^{-1}$, the instability is highly non-linear and the spectrum becomes a lot smoother as the wavenumber fluctuations are no longer independent.

On Figure \ref{fig:FFT_By_t}(b), we show spatially averaged (indicated by $\langle ... \rangle$) variations of different vector parameters from their value at $t = 0$.
The variation of a vector $\bm{a}$ is defined as $\langle | \delta \bm{a} | \rangle = \sqrt{\Delta a_x^2 + \Delta a_y^2 + \Delta a_z^2}$ with $\Delta a_i (t) = a_i (t) - a_i (t=0)$ for each component $i = x,y,z$.
During the linear stage ($t < 400 \, \Omega_{ci}^{-1}$), the magnetic field variations $\langle | \delta \bm{B} | \rangle$ undergo a relatively slow increase, which then abruptly amplify when the non-linear stage is reached at $t = 400 \, \Omega_{ci}^{-1}$.
The same behavior is observed, albeit with smaller amplitudes, for both the plasma bulk speed $\langle | \delta \bm{u} | \rangle$ and electric current density $\langle | \delta \bm{j} | \rangle$ variations.
These changes are due to the triggering of more violent reconnection events, which leads to the emergence of large scale structures during the non-linear stage of the instability.

\begin{figure}[t!] 
    \resizebox{\hsize}{!}{\includegraphics{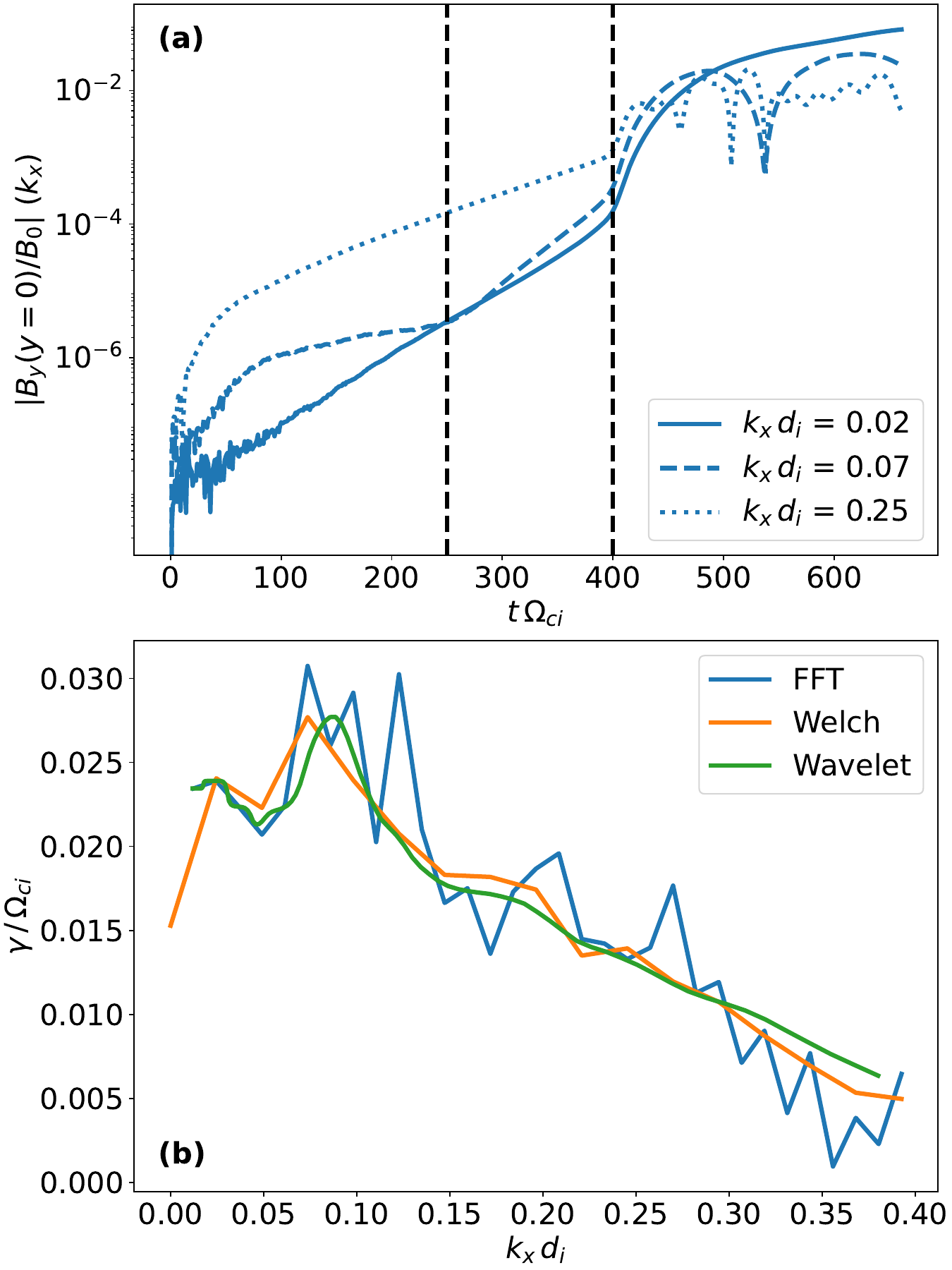}}
    \caption{\label{fig:linear_growth_rates}
        Linear growth rates of the tearing instability and examples of some associated magnetic perturbations' evolution.
        Panel (a) provides the temporal evolution of $|B_y (y=0)| (k_x)$ (decomposed using a FFT) for three wavenumbers $k_x \, d_i = 0.02$ (plain curve), $k_x \, d_i = 0.07$ (dashed curve) and $k_x \, d_i = 0.25$ (dotted curve).
        The two black vertical dashed lines indicate the time interval considered for the linear growth rates estimation $\gamma$, between $t = 250 \, \Omega_{ci}^{-1}$ and $t = 400 \, \Omega_{ci}^{-1}$.
        Panel (b) gives $\gamma / \Omega_{ci} (k_x \, d_i)$, estimated from $|B_y (y=0)| (k_x)$ decomposed using: a regular FFT (blue), the Welch method (orange) and wavelets (green).
    }
\end{figure}  

Figure \ref{fig:linear_growth_rates}(a) gives the temporal evolution of $|B_y (y=0)| (k_x)$, decomposed over wavenumbers using an FFT, for three different modes: $k_x \, d_i = 0.02$, $k_x \, d_i = 0.07$, and $k_x \, d_i = 0.25$.
The fluctuations are significantly affected by the noise for $t \lesssim 75 \, \Omega_{ci}^{-1}$, after which the instability dominates and fluctuations follow an exponential growth until the beginning of the non-linear stage at $t = 400 \, \Omega_{ci}^{-1}$.
We note that fluctuations at very low wavenumbers ($k_x \, d_i = 0.02$) continue to grow during the non-linear stage, as they are related to the apparition and development of large-scale magnetic islands, which we discuss with more details in the next section. 
The slope (in log scale) of $|B_y (y=0)| (k_x \, d_i = 0.07)$ (dashed curve) moreover changes around $t = 250 \, \Omega_{ci}^{-1}$, so in the following we only consider later times ($t > 250 \, \Omega_{ci}^{-1}$) for the growth rates estimation.

The dispersion relation of the tearing instability linear growth rates $\gamma (k_x)$ is shown in Figure \ref{fig:linear_growth_rates}(b).
Those were obtained by fitting of $B_y (y=0, k_x, t) \propto e^{\gamma(k_x)}$ for each mode $k_x \, d_i < 0.4$ along the current sheet in the time range $t \in [250, 400] \, \Omega_{ci}^{-1}$, between the black vertical dashed lines in Figure \ref{fig:linear_growth_rates}(a).
As mentioned earlier, a few modes (those around $k_x \, d_i = 0.07$) do not exhibit an exponential increase with a constant $\gamma$, so the fitting rather corresponds to the growth rates at the end of the linear stage.
Colored curves correspond to growth rates fittings using three different decomposition of the perturbations $B_y (y=0)$ over $k_x$.
The blue curve uses a regular FFT, the orange one the Welch method \citep{Welch_1967_FFT} with a Hann windowing of two segments at 50 \% overlap \citep{Blackman-Tukey_1958_power_spectra_Hann_window}, and the green curve a wavelet transform with a Morlet mother function of normalized angular frequency $\omega_0 = 6$ \citep{Torrence-Compo_1998_wavelet}.
The growth rate attains its maximum $\gamma \sim 0.25 \, \text{-} \, 0.30 \, \Omega_{ci}$ at $k_x d_i = 0.07$ (0.09 for the wavelet transform) and steadily decreases for larger values of $k_x \, d_i$.
The fluctuations at wavenumbers with the highest growth rates are however not necessarily the most energetics at the end of the linear phase.
Indeed, the perturbation amplitudes at the end of the linear stage also depends on their initial values after the instability emergence from the noise ($t \sim 100 \Omega_{ci}$), which can vary by two orders of magnitude (see Figure \ref{fig:linear_growth_rates}(a)).
Moreover, since the perturbations at wavenumbers around $k_x d_i = 0.07$ have a slower increase at the beginning (for $t \in [100, 250] \, \Omega_{ci}^{-1}$) than for $t \in [250, 400] \, \Omega_{ci}^{-1}$, as can be seen on Figure \ref{fig:FFT_By_t}(a) and Figure \ref{fig:linear_growth_rates}(a), their growth rates plotted on Figure \ref{fig:linear_growth_rates}(b) are not representative of the entire linear stage.
In any case, the instability becomes fully non-linear after $t = 400 \, \Omega_{ci}^{-1}$, and it becomes necessary to investigate the system's behavior in real space. 

\subsection{Non-Linear Stage and Magnetic Islands Dynamics}

\begin{figure*}[t!] 
    \resizebox{\hsize}{!}{\includegraphics{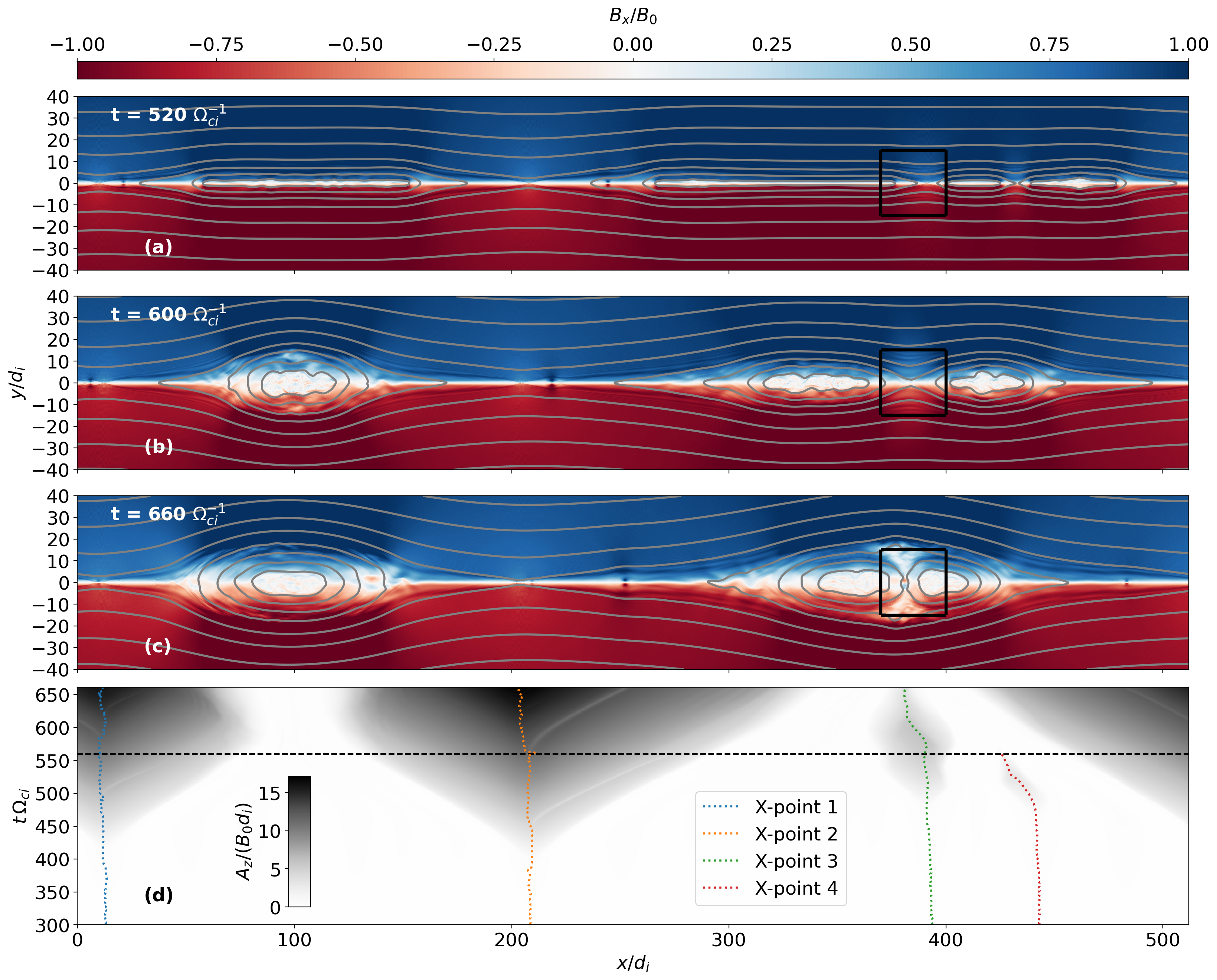}}
    \caption{\label{fig:NL_evolution_whole-box}
        Global magnetic field evolution during the non-linear stage of the tearing instability.
        Panels (a), (b) and (c) display, in color, the magnetic field component $B_x / B_0$ for a zoom around $y \in [-40, 40] \, d_i$, and at times $t = 520 \, \Omega_{ci}^{-1}$, $t = 600 \, \Omega_{ci}^{-1}$ and $t = 660 \, \Omega_{ci}^{-1}$, respectively.
        Grey curves are magnetic field lines, defined as isovalues of the magnetic vector potential's out-of-plane component $A_z$.
        Panel (d) shows $A_z (x, y=0, t)$: the temporal evolution of $A_z$ in the center of the current sheet.
        Positions of the principal X-points, defined as local maxima of $A_z (y=0)$, are indicated by the differently colored dotted curves.
    }
\end{figure*}  

Figure \ref{fig:NL_evolution_whole-box} summarizes the evolution of the magnetic configuration during the non-linear stage of the instability (starting at $t = 400 \, \Omega_{ci}^{-1}$).
The normalized magnetic field reconnecting component $B_x / B_0$ is displayed in panels (a-c) for three different times $t = 520, \, 600$, and $660 \, \Omega_{ci}^{-1}$, within $y \in [-40, 40] \, d_i$.
We observe classical signatures of the non-linear tearing instability.
Large-scale magnetic islands emerge between the different principal X-points.
These islands initially contract along $x$ and become wider along $y$.
Then, they grow in both directions once their contraction has stopped and they continue accumulating reconnected plasma, see e.g., the leftmost island in Figure \ref{fig:NL_evolution_whole-box} (a-c).
Some islands undergo coalescence, as can be seen with the three rightmost islands in panel (a) becoming two islands in panel (b), and then beginning to merge together in panel (c).
We moreover observe smaller islands being sporadically ejected with the outflows from the main X-points.
These will eventually reach and coalesce with the larger islands.

The temporal evolution of $A_z (y=0, x, t)$, the out-of-plane vector potential along the center of the current sheet, is shown in Figure \ref{fig:NL_evolution_whole-box}(d).
Local maxima of $A_z (y=0, x, t)$ are used to define the principal X-points (labeled 1-4) positions at each time, as indicated by the colored curves, \citep[see also][]{Markidis_2012_plasmoid_chain_coalescence}.
When the non-linear stage is reached, there is an important increase of $A_z$ spreading around the X-points as the reconnected field lines are being advected by the outflows.
Thin whiter lines within the outflow regions (e.g. on the right of X-point 2) are associated with the formation and ejection of smaller plasmoids produced by secondary reconnection events near the principal X-points.
Coalescences events are marked by diminishing values of $A_z$ around the X-point 3, as well as the X-point 4, which disappears after $t \simeq 560 \, \Omega_{ci}^{-1}$ (black horizontal dashed line) when the two surrounding islands have nearly fully merged.

\begin{figure*}[t!] 
    \resizebox{\hsize}{!}{\includegraphics{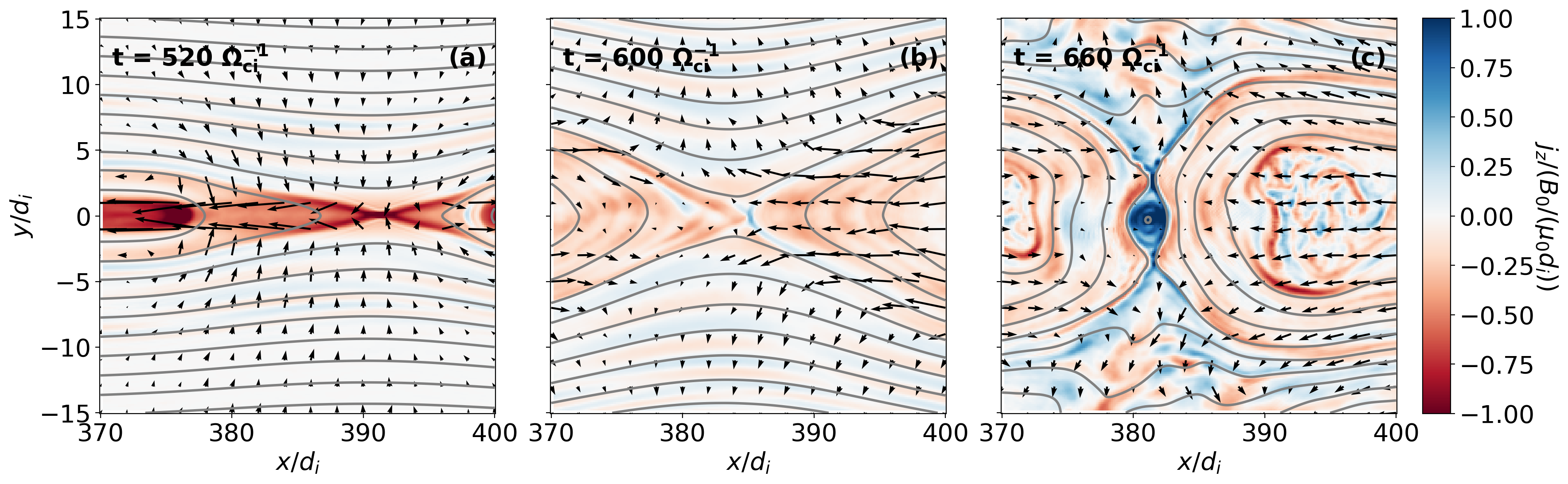}}
    \caption{\label{fig:NL_evolution_sub-box}
        Transition of a primary reconnection X-point into a coalescence site (also reconnecting) between two merging magnetic islands.
        The three panels present the out-of-plane electric current density $j_z$ within the black sub-box of Figure \ref{fig:NL_evolution_whole-box} (a-c), and for the three same times $t = 520, \, 600$, and $660 \, \Omega_{ci}^{-1}$.
        The $j_z$ values have been smoothed by taking a spatial average of total lengths $d_i/2$ in the $x$ and $y$ directions around each cell.
        }
\end{figure*}  

Let us consider in more details X-point~3 and the coalescence process between the two surrounding plasmoids.
Figure \ref{fig:NL_evolution_sub-box} shows the out-of-plane electrical current density $j_z$ (colors) and ion plasma outflow direction (arrows) for a zoom at $x = [370, 400] \, d_i$, $y = [-15,15] \,  d_i$ and for times 520~(a), 600~(b) and 660~$\Omega_{ci}^{-1}$~(c) , see black sub-box in Figure \ref{fig:NL_evolution_whole-box}(a-c).
At $t = 520 \, \Omega_{ci}^{-1}$ (panel (a)), there is still a primary reconnection event with $j_z < 0$ around the X-point, as well as inflows from the $\pm \, y$ directions and outflows in the $\pm \, x$ directions.
Then, due to the large-scale dynamics of the original current sheet, plasmoids from the two sides of the X-point get closer to one another, and their outer layer's magnetic field start to reconnect, see panel (b) where $t = 600 \, \Omega_{ci}^{-1}$.
This leads to a gradual reversal of the reconnection inflow-outflow directions and of the electrical current $j_z$ component.
As reconnection proceeds, we even observe the formation of a smaller magnetic island (panel (c)), with an inverted current system ($j_z > 0$) as compared to the primary plasmoids.

\subsection{Evaluation of the Reconnection Rates}

\begin{figure}[t!] 
    \resizebox{\hsize}{!}{\includegraphics{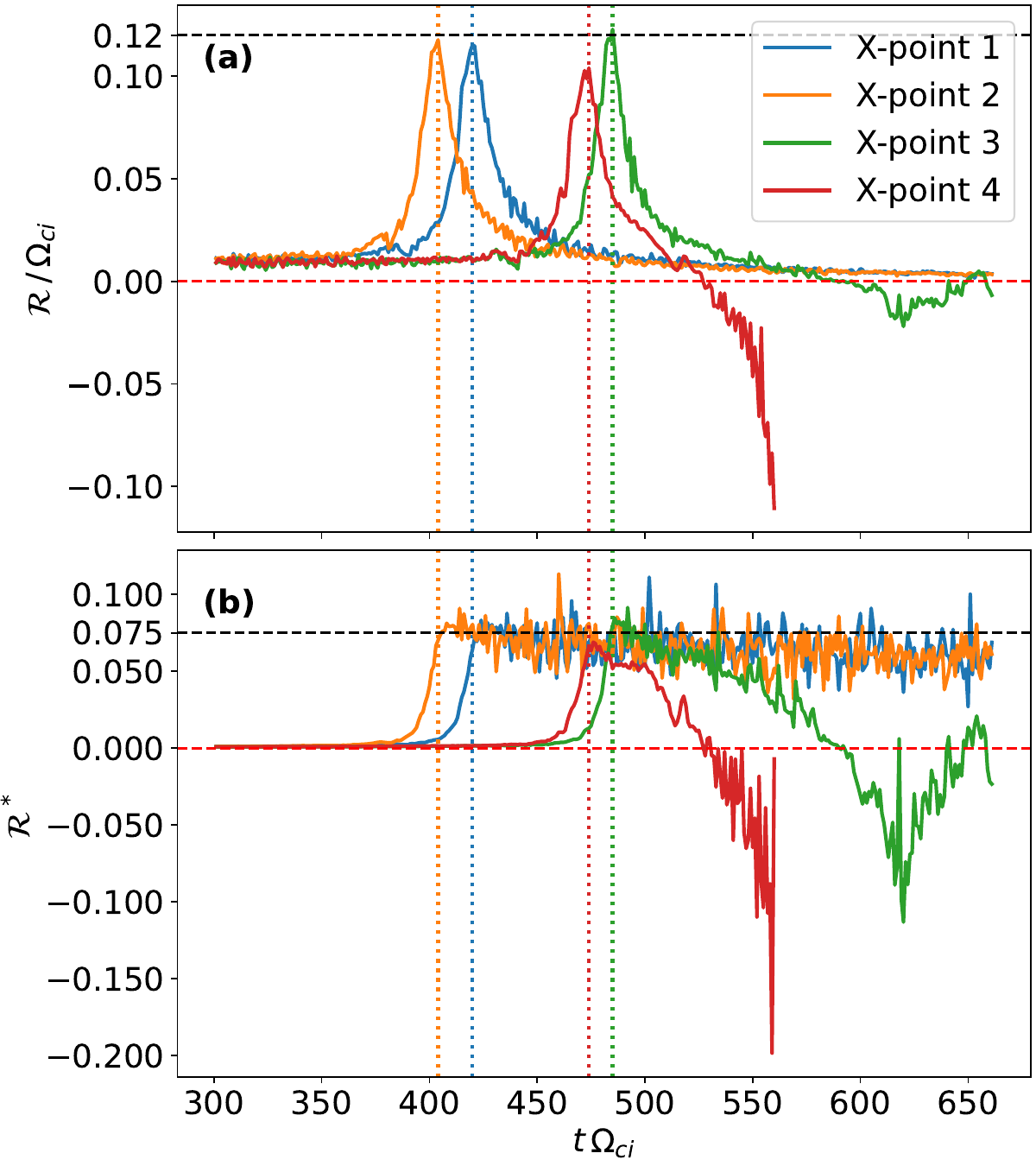}}
    \caption{\label{fig:rec_rate}
    Temporal evolution of the reconnection rates for the different X-points, with same associated colors as in Figure \ref{fig:NL_evolution_whole-box} (d).
    Panel (a) and (b) show the reconnection rates $\mathcal{R}$ and $\mathcal{R}^*$, defined in Equations (\ref{eq:reconnection_rate_flux}) and (\ref{eq:reconnection_rate_elec}), respectively.
    The colored vertical dashed lines give the times at which the corresponding $\mathcal{R}$ are at their maximum, indicating that the non-linear reconnection regime is reached.
    }
\end{figure}  

In order to quantify the efficiency of magnetic reconnection, let us now consider the reconnection rates evolution for the principal X-points identified in Figure \ref{fig:NL_evolution_whole-box}.
More precisely, we estimate two different reconnection rates proxies.

We first consider the relative magnetic flux transfer rate \citep[see][]{Papini_2019_Hall-MHD_turbulence}, defined for each X-point as:
\begin{equation}
    \mathcal{R} = \frac{1}{\Phi (t)} \frac{\partial \Phi (t)}{\partial t} \label{eq:reconnection_rate_flux}
\end{equation}
where $\Phi(t) = A_z^X (t) - A_z^O (t)$ with $A_z^X$ and $A_z^O$ the out-of-plane vector potential at an X-point and O-point, respectively.
We estimate $A_z^X$ as an average of $A_z$ around its local maxima for $\pm \, d_i/2$ along $x$, and we considered $A_z^O = 0 \, \forall t$ as the $A_z$ values around the O-points are negligible.

Figure \ref{fig:rec_rate}(a) shows the temporal evolution of $\mathcal{R}$ for the X-points 1-4, which all share common features.
\begin{enumerate}
    \item The X-points first have a same constant $\mathcal{R} \sim 0.01 \, \Omega_{ci}$ during the linear stage, implying an exponential transfer of magnetic flux.
    \item Then, each of them undergo an increase, going from $\mathcal{R} \sim 0.01 \, \Omega_{ci}$ to a maximum of $\mathcal{R} \sim 0.12 \, \Omega_{ci}$ in about $50 \, \Omega_{ci}^{-1}$.
    \item Finally, the reconnection rates $\mathcal{R}$ drop, going back to $\mathcal{R} \sim 0.01 \, \Omega_{ci}$ in $\sim 75 \, \Omega_{ci}^{-1}$, and slowly decrease towards zero over time.
    This decrease is mainly due to the normalization by $1/\Phi (t)$ since $\partial_t \Phi >0$.
\end{enumerate}
Even though they are similar, all these profiles are shifted from one another, meaning that the non-linear stage is reached at different times for each X-point.
X-points 3 and 4 eventually reach negative values of $\mathcal{R}$ and $\mathcal{R}^*$ because of the coalescence process and associated reversal of magnetic flux transfer.
A slightly lower maximum value of $\mathcal{R} \sim 0.1 \, \Omega_{ci}$ is moreover attained for X-point~3.
This is due to the surrounding islands starting to merge quickly after the onset of the X-point's non-linear stage.

The second reconnection rate proxy we consider is
\begin{equation}
    \mathcal{R}^* = - \frac{E_z^X}{v_{A,0} B_0} \label{eq:reconnection_rate_elec}
\end{equation}
with $E_z^X$ the out-of-plane electric field at the X-point, and $v_{A,0}$ the initial background Alfv\'en speed.

Figure \ref{fig:rec_rate}(b) displays the evolution of $\mathcal{R}^*$, which, as for $\mathcal{R}$, have very similar profiles for the four X-points, albeit shifted in time from one another.
Values of $\mathcal{R}^*$ are first nearly zero ($\sim 10^{-3}$) and then present an abrupt increase (with timescale $25 \, \Omega_{ci}^{-1}$) as they transition into the non-linear stage.
Each X-point attains a maximum of $\mathcal{R}^* \sim 0.075$, in agreement with the seemingly universal value of order 0.1 in the context of weakly collisional plasmas \citep[see][and references therein]{Cassak_2017_review_01_rec_rate}.
Reconnection is continuously operating at X-points 1 and 2 as indicated by the values of $\mathcal{R}^*$ staying of the same order, and only slowly decrease till reaching $\sim 0.06$ at the end of the simulation.
For X-points 3 and 4, $\mathcal{R}^*$ decreases faster after having reached its maximum, and eventually becomes negative (as for $\mathcal{R}$), because of the coalescence process.

The differences in the $R$ and $R^*$ profiles is due to their normalizations: $R$ is normalized by $\Phi (t)$ which is temporally increasing, while $R^*$ is normalized by a constant $1/(v_{A,0} B_0)$.
Both definitions have their advantages.
From $R$, we clearly see that there is a transfer of magnetic flux, hence reconnection, even during the linear phase.
Considering, as a first approximation, $R$ to be nearly constant during this first phase implies a quasi-exponential growth of the reconnected flux $\Phi(t) \propto e^{Rt}$.
This would be expected from a linear instability with one largely dominant mode  $\gamma_m = R$ \citep[see][]{Betar_2022_linear_tearing_tutorial}.
In our case, as shown in Figures (\ref{fig:FFT_By_t}) and (\ref{fig:linear_growth_rates}), a range of modes is excited, but during the linear stage, the fluxes at the X-points shown in Figure \ref{fig:NL_evolution_whole-box}(d) are still exponentially growing.
From $R^*$, we see that in the non-linear phase, the purely reconnecting X-points 1 \& 2 reach a quasi-steady state regime of "fast reconnection" with values of the order $R^* \sim 0.1$, with a nearly constant non-ideal out-of-plane electric field Ez (and thus transfer of flux $\partial_t \Phi$).
Then, the definitions of $R$ and $R^*$ seem more adapted for study of the linear and non-linear stage, respectively.

During the linear stage, the X-points within the current sheet are more numerous (not shown).
The X-points with the largest $A_z$ values are also the one attaining the non-linear stage.
This is why we can track their position during the linear stage by considering local maxima of $A_z$, see Figure \ref{fig:NL_evolution_whole-box}(d).

For each of the X-points 1-4, we consider that the non-linear reconnection regime is attained when the reconnection rate $\mathcal{R}$ is at its maximum, also corresponding to when $\mathcal{R}^*$ reaches its value of order 0.1.
These times, we denote $t_{NL}$ in the following, are marked by the colored vertical dashed lines in Figure \ref{fig:rec_rate}.
We observe that, when reaching the non-linear stage, each X-points has nearly the same vector potential value $A_z (y= 0, t = t_{NL}) \simeq 0.8 \, B_0 d_i$.
We then consider that the different X-points transition towards the non-linear regime last for a common duration of $\tau_{NL} \sim 25 \text{-} 50 \, \Omega_{ci}^{-1}$ because the $\mathcal{R}$ profiles are so similar.
This thus defines a critical value of the vector potential, $A_{z,c} = A_z (y=0, t_{NL} - \tau_{NL}) \simeq 0.3 \text{-} 0.4  \, B_0 d_i$, after which the non-linear regime is triggered.

\section{Conversion and Evolution of Energy} \label{sec:nrj_conv}

\subsection{Energy Conversion and Pressure-Strain Interaction}

For a collisionless plasma constituted of several species $s$, with associated mass density $\rho_s$, velocity $\bm{u}_s$, pressure tensor $\mathbf{P}_s$, heat flux $\bm{q}_s$ and electrical current density $\bm{j}_s$, the energy evolution equations write:
\begin{align}
    \partial_t \left( \frac{B^2}{2 \mu_0} + \frac{\varepsilon_0 E^2}{2} \right) &= - \bm{j} \cdot \bm{E} -  \bm{\nabla} \cdot \left( \frac{\bm{E} \times \bm{B}}{\mu_0} \right) \label{eq:EM_nrj_conv} \\
    \partial_t \left( \frac{\rho_s u_s^2}{2} \right) &= \mathbf{P}_s : \bm{\nabla u}_s + \bm{j}_s \cdot \bm{E} \nonumber \\ & - \bm{\nabla} \cdot \left( \frac{\rho_s u_s^2}{2} \bm{u}_s + \mathbf{P}_s \cdot \bm{u}_s \right) \label{eq:kin_nrj_conv} \\
    \partial_t \left( \frac{3}{2} P_s \right)  &= - \mathbf{P}_s : \bm{\nabla u}_s \nonumber \\ & - \bm{\nabla} \cdot \left( \frac{3}{2} P_s \bm{u}_s + \bm{q}_{s} \right), \label{eq:thermal_nrj_conv}
\end{align}
where $\bm{j} = \sum_s \bm{j}_s$ and $P_s = \text{tr} (\mathbf{P}_s) /3$.
Equation (\ref{eq:EM_nrj_conv}) is the Poynting's theorem, while Equations (\ref{eq:kin_nrj_conv}) and (\ref{eq:thermal_nrj_conv}) result from manipulation of the Vlasov's equation integrated moments (from order 0 to 2), with the different fluid quantities defined as integrated moments of the particles velocity distributions functions \citep{Braginskii_1958, Braginskii_1965}.
It is apparent from Equations (\ref{eq:EM_nrj_conv})-(\ref{eq:thermal_nrj_conv}) that locally, the temporal evolution of an energy type, whether electromagnetic, kinetic, or internal, has two contributions.
One is due to the divergence of different fluxes, redistributing the energy spatially within the medium.
The other comes from the electrical work rate ($\bm{j}_s \cdot \bm{E}$) and pressure-strain interaction ($- \mathbf{P}_s : \bm{\nabla u}_s$) terms, converting one type of energy into another.

Spatial averages of Equations (\ref{eq:EM_nrj_conv})-(\ref{eq:thermal_nrj_conv}) over a domain with periodic boundary conditions \citep{Yang_2017_pressure-strain} give
\begin{align}
    \partial_t \langle \frac{B^2}{2 \mu_0} + \frac{\varepsilon_0 E^2}{2} \rangle &= - \langle \bm{j} \cdot \bm{E} \rangle, \label{eq:EM_nrj_conv_avg} \\
    \partial_t \langle \frac{\rho_s u_s^2}{2} \rangle &= \langle \mathbf{P}_s : \bm{\nabla u}_s \rangle + \langle \bm{j}_s \cdot \bm{E} \rangle, \label{eq:kin_nrj_conv_avg} \\
    \partial_t \langle \frac{3}{2} P_s \rangle &= - \langle \mathbf{P}_s : \bm{\nabla u}_s \rangle, \label{eq:thermal_nrj_conv_avg}
\end{align}
and there are no contributions from the divergence of fluxes anymore since the domain is closed.
The average energy densities temporal evolution are thus directly related to the electrical work rates $\bm{j}_s \cdot \bm{E}$ and pressure-strain interaction terms $\mathbf{P}_s : \bm{\nabla u}_s$.
We note that the pressure-strain term is often further decomposed into several parts as to get better insights on the processes related to conversion towards internal energy \citep{Del-Sarto_2016_anisotropy_velocity_shear,Yang_2017_pressure-strain,Zhou_2021_agyrotropy_pressure-strain_MMS,Hellinger_2025_PS_agyrotropy}.
In the following, we however prefer to keep the pressure-strain term as it is and thus only consider the total conversion of energy going towards ion heating for simplicity.
Moreover, in the studied simulation, the electrons (subscript $e$) are massless and isothermal, so they do not contribute to the global energy budget evolution as seen from Equations (\ref{eq:kin_nrj_conv_avg}) and (\ref{eq:thermal_nrj_conv_avg}): $- \langle \mathbf{P}_e : \bm{\nabla u}_e \rangle =0$ and $\langle \bm{j}_e \cdot \bm{E} \rangle = 0$.

It can be shown that Equations (\ref{eq:EM_nrj_conv_avg})-(\ref{eq:thermal_nrj_conv_avg}) remain valid in the case of M\"obius periodic conditions used in the present simulation.
Considering a regular periodic domain of volume $\mathcal{V}$ encompassed in a closed surface $\Sigma$, the Green-Ostrogradski theorem writes
\begin{equation}
    \int_\mathcal{V} (\bm{\nabla} \cdot \bm{a}) \, dV = \oint_\Sigma \bm{a} \cdot d \bm{S} = 0 \label{eq:Green-Ostrogradski}
\end{equation}
for any vector $\bm{a}$.
This is also valid for M\"obius boundaries, as: only the components normal to $d \bm{S}$ change sign so they do not influence the scalar product, and the change of spatial coordinate does not affects the result because of the integration.
Although strictly speaking the M\"obius strip is a non-orientable surface, the simulation domain remains a rectangle box with boundary properties akin to a M\"obius strip, so Equation (\ref{eq:Green-Ostrogradski}) is still applicable.

\subsection{Global Energy Conversion and Heating during the Tearing Instability} \label{sec:nrj_conv_global}

\begin{figure}[t!] 
    \resizebox{\hsize}{!}{\includegraphics{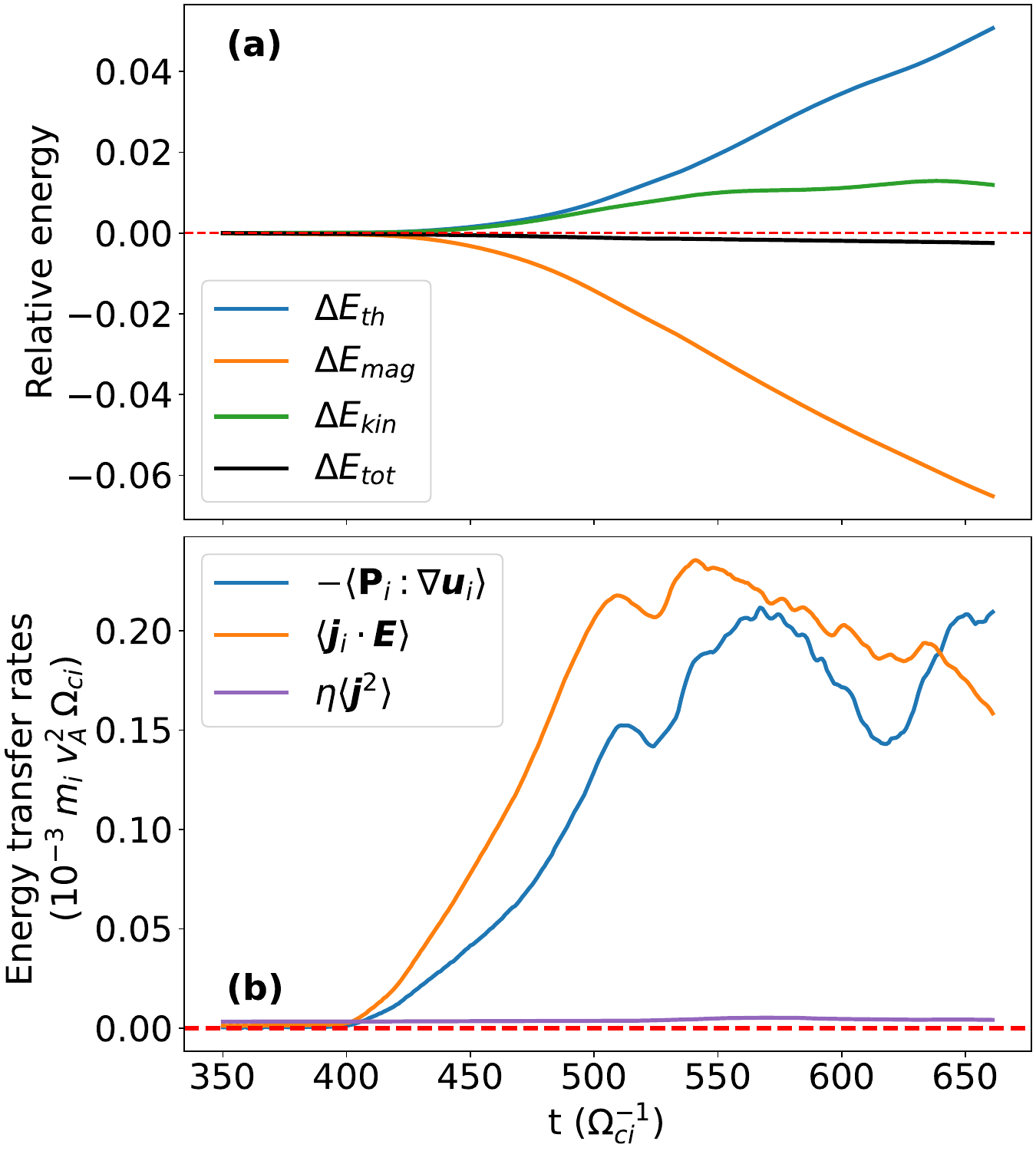}}
    \caption{\label{fig:nrj_PS_etc}
    Global energy conversion and evolution during the tearing instability.    
    Panel (a) shows the temporal evolution for different type of energies $\Delta E(t)$ averaged over the domain.
    Panel (b) gives the related energy conversion terms $\langle \bm{j}_i \cdot \bm{E} \rangle$ (orange curve) and $\langle \mathbf{P}_i : \bm{\nabla u}_i \rangle$ (blue curve) as defined by Equations (\ref{eq:EM_nrj_conv_avg}), (\ref{eq:kin_nrj_conv_avg}) and (\ref{eq:thermal_nrj_conv_avg}).
    }
\end{figure}  

We focus here on the global energy conversion and evolution during the tearing instability.
The temporal evolution of several relative energy densities, normalized and spatially averaged over the domain, are displayed in Figure \ref{fig:nrj_PS_etc}(a).
The averaged energy densities we consider are: the ion (with subscript $i$) internal energy $E_{th} = 3 \langle P_i \rangle / 2$, the ion bulk kinetic energy $E_{kin} = \langle \rho_i u_i^2 \rangle / 2$, the magnetic energy $E_{mag} = \langle B^2 \rangle / (2 \mu_0)$ and the total energy $E_{tot} = E_{th} + E_{kin} + E_{mag}$.
For each spatially averaged energy density $E (t)$, we define $\Delta E (t) = (E(t) - E(t=0)) / E_{tot} (t=0)$.
In the simulation, the electromagnetic energy is equal to the magnetic energy.
Figure \ref{fig:nrj_PS_etc}(b) shows the spatially averaged ion pressure-strain interaction $\langle \mathbf{P}_i : \bm{\nabla u}_i \rangle$ and electrical work rate $\langle \bm{j}_i \cdot \bm{E} \rangle$, corresponding to the energy conversion rates as described by Equations (\ref{eq:EM_nrj_conv_avg} - \ref{eq:thermal_nrj_conv_avg}), as well as the resistive electrical work rate $\eta \langle \bm{j}^2 \rangle$.
Here, and in the following of the study, spatial derivatives ($\bm{\nabla}$) are evaluated through a centered finite difference scheme.

It is clear on Figure \ref{fig:nrj_PS_etc} that most of the energy conversion takes place during the non-linear stage of the instability.
Indeed, starting around $t = 400 \, \Omega_{ci}^{-1}$, there is, at all times, an important decrease of magnetic energy $\Delta E_{mag}$ (orange curve) while the ion internal energy $\Delta E_{th}$ (blue) increases.
At the beginning of the non-linear stage, $\Delta E_{th}$ and $\Delta E_{kin}$ (green) are nearly equal, as the same amount of energy is supplied into ion heating and bulk acceleration.
After $t = 540 \, \Omega_{ci}^{-1}$, there is a quasi-saturation of $\Delta E_{kin}$ while $\Delta E_{th}$ continues to significantly increase.

The temporal evolutions of $\langle \bm{j}_i \cdot \bm{E} \rangle$ and $- \langle \mathbf{P}_i : \bm{\nabla u}_i \rangle$ present an important increase after $t = 400 \, \Omega_{ci}^{-1}$ (beginning of the non-linear stage), see Figure \ref{fig:nrj_PS_etc}(b).
Since $\langle \bm{j}_i \cdot \bm{E} \rangle > 0$, there is a transfer of magnetic energy towards the ions, while $- \langle \mathbf{P}_i : \bm{\nabla u}_i \rangle >0$ indicates ion heating.
These energy conversion rates profiles are moreover very similar until $t = 540 \, \Omega_{ci}^{-1}$, albeit with different amplitudes ($- \langle \mathbf{P}_i : \bm{\nabla u}_i \rangle / \langle \bm{j}_i \cdot \bm{E} \rangle \in [0.5, 0.75]$).
After $t = 540 \, \Omega_{ci}^{-1}$, $\langle \bm{j}_i \cdot \bm{E} \rangle$ progressively decreases as reconnection slows down at the different X-points.
The $- \langle \mathbf{P}_i : \bm{\nabla u}_i \rangle$ term decreases after $t = 570 \, \Omega_{ci}^{-1}$, but it then re-increases around $t = 620 \, \Omega_{ci}^{-1}$, which seems to be due to the coalescence of two magnetic islands around X-point 3, see the next section for more details.

During the linear stage, there is a dominance of the resistive term $\eta \langle \bm{j}^2 \rangle$ which is about 5-10 times higher than $- \langle \mathbf{P}_i : \bm{\nabla u}_i \rangle$ and 2-3 times higher than $\langle \bm{j}_i \cdot \bm{E} \rangle$.
However, during the non-linear stage, where the quasi-totality of the energy conversion happens, $\eta \langle \bm{j}^2 \rangle$ rapidly becomes negligible as the other terms get 30-50 times higher than it.
Thus, in this simulation, the transition from linear to non-linear stage also correspond to the reconnection regime shifting from mostly resistive to nearly collisionless.
This is consistent with Figure \ref{fig:rec_rate}(b), where the normalized reconnection rates values $\sim 0.1$, typical of collisionless reconnection, are observed only after the instability non-linear stage is reached.

The energy conversion is however non-uniform, so in the next section we consider its local properties and evolution in different parts of the reconnecting current sheet.

\subsection{Local Energy Conversion} \label{sec:nrj_conv_local}

\begin{figure*}[t!] 
    \resizebox{\hsize}{!}{\includegraphics{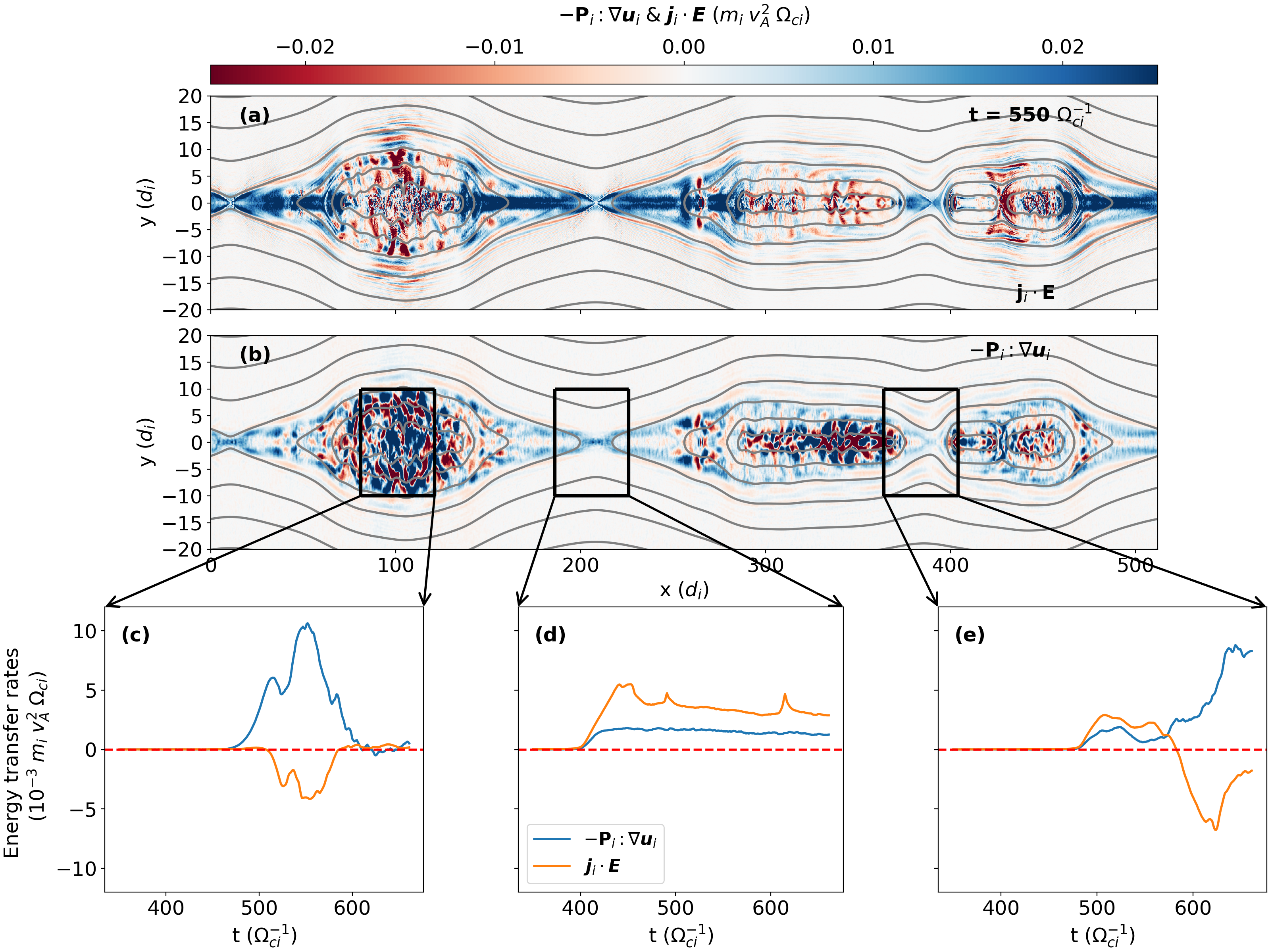}}
    \caption{\label{fig:local_nrj_transfer}
        Local energy conversion rates during the non-linear stage of the tearing instability.
        Panels (a) and (b) show the spatial distribution of $\bm{j}_i \cdot \bm{E}$ and $- \mathbf{P}_i : \bm{\nabla u}_i$, at time $t = 550 \, \Omega_{ci}^{-1}$ and for $y \in [-20, 20] \, d_i$.
        The pressure-strain interaction term has been locally averaged over a window of total length $2 \, d_i$ in the $x$ and $y$ directions to reduce the noise.
        Colorbar values of the two panels have moreover been constrained to $[-2.5, 2.5] \times 10^{-2} \, m_i v_A^2 \Omega_{ci}$.
        Panels (c), (d) and (e) give the temporal evolution of $- \mathbf{P}_i : \bm{\nabla u}_i$ (blue curves) and $\bm{j}_i \cdot \bm{E}$ (orange curves) spatially averaged over the black rectangle sub-boxes in panel (b).
        }
\end{figure*}  

In this section, we investigate the local energy conversion during the non-linear stage of the tearing instability.
The instantaneous spatial distribution (at time $t = 550 \, \Omega_{ci}^{-1}$ and for $y \in [-20, 20] \, d_i$) of the ion electrical work rate $\bm{j}_i \cdot \bm{E}$ and pressure-strain interaction $- \mathbf{P}_i : \bm{\nabla u}_i$ are presented in Figure \ref{fig:local_nrj_transfer}(a) and (b), respectively.
These two energy conversion rates are highly inhomogeneous within the reconnecting current sheet.
Both $\bm{j}_i \cdot \bm{E}$ and $- \mathbf{P}_i : \bm{\nabla u}_i$ are positive around the X-points, indicating magnetic energy transfer and heating towards ions.
All along the current sheet, and outside the magnetic islands, there is $\bm{j}_i \cdot \bm{E} > 0$, which is consistent with Fermi acceleration of the ions by the newly reconnected field lines \citep{Drake_2010_proton_Fermi_acc_firehose}.
The situation is more complex within the magnetic islands where positive and negative patches are present for the two energy conversion rates.

To better quantify the local energy conversion rates and their evolution, we considered spatial averages of $\bm{j}_i \cdot \bm{E}$ and $- \mathbf{P}_i : \bm{\nabla u}_i$ around three different parts of the reconnecting current sheet: the center of a magnetic island, a primary X-point, and another X-point between two coalescing islands.
These averages are displayed in Figure \ref{fig:local_nrj_transfer}(c,d,e), and evaluated within the three rectangle sub-boxes indicated by black plain lines in Figure \ref{fig:local_nrj_transfer}(b).
The sub-boxes limits are all defined with $y \in [-10, 10] \, d_i$, and then $x \in [81, 121] \, d_i$ for the magnetic island, $x \in [186, 226] \, d_i$ for the primary X-point and $x \in [364, 404] \, d_i$ for the coalescing islands.
We chose the three sub-boxes to have the same dimensions (of $40 \times 20 \, d_i$) to allow direct comparison between the local averages.
These sub-boxes are not closed domains, so the local energy densities inside also vary due to the non-zero divergence of fluxes, see Equations (\ref{eq:EM_nrj_conv}), (\ref{eq:kin_nrj_conv}) and (\ref{eq:thermal_nrj_conv}).
However, the electrical work rate and pressure-strain interaction terms still quantify the conversion from one type of energy to another.
Investigating the evolution of local fluxes and associated divergence would be useful to get better insights about where the energy then gets redistributed within the plasma, but goes out of the scope of the present study.

Figure \ref{fig:local_nrj_transfer}(d) shows the energy conversion rates temporal evolution around the primary X-point 2, which always exhibits steady reconnection (no coalescence is observed) as can be seen in Figures \ref{fig:NL_evolution_whole-box} and \ref{fig:rec_rate}.
As for the global rates discussed in Section \ref{sec:nrj_conv_global}, the conversion terms around the X-point are negligible before $t = 400 \, \Omega_{ci}^{-1}$ and increase rapidly after this time, which corresponds to the beginning of non-linear stage and collisionless reconnection regime.
Once their maximum is attained, around $t \simeq 450 \, \Omega_{ci}^{-1}$, the conversion rates slowly decrease during the rest of the simulation (by $\sim 25 \%$ at the end of it), as expected from the same behavior of the reconnection rate in Figure \ref{fig:rec_rate}(b).
Moreover, there are smaller spikes in $\bm{j}_i \cdot \bm{E}$ (orange curve), which are linked with secondary plasmoids' ejection, and are however not present in pressure-strain interaction term (blue curve).
Once the non-linear stage is attained, there is on average less heating than energy going towards bulk speed acceleration as $- \mathbf{P}_i : \bm{\nabla u}_i /( \mathbf{P}_i : \bm{\nabla u}_i + \bm{j}_i \cdot \bm{E} ) \sim 0.80$.
This however depends on the sub-box size, and taking instead sub-boxes of dimensions $20 \times 10 \, d_i$ and $10 \times 5 \, d_i$ (not shown) yield ratio of $- \mathbf{P}_i : \bm{\nabla u}_i / ( \mathbf{P}_i : \bm{\nabla u}_i + \bm{j}_i \cdot \bm{E} )$ closer to $1.15$ and $1.35$, respectively, implying more heating than conversion towards ion bulk speed in the close vicinity of the X-point.

Temporal evolution of the averaged energy conversion rates around the center of a magnetic island are shown in Figure \ref{fig:local_nrj_transfer}(c).
The electrical work rate is negative $\bm{j}_i \cdot \bm{E} < 0$, thus, some energy is converted from the ions towards the magnetic field.
The ion pressure-strain is positive, $- \mathbf{P}_i : \bm{\nabla u}_i > 0$, which indicates a conversion towards increasing internal energy.
For $t < 450 \, \Omega_{ci}^{-1}$, the energy conversion terms are zero because the jets have not yet reached the sub-box.
Then, as the jets collide and the island forms, the amplitude of both terms increase.
As the island grows, the terms amplitude reach a maximum at $t = 550 \, \Omega_{ci}^{-1}$, and finally decrease until reaching zero again.
This decrease is due to the plasmoid getting to a size such that the newly reconnected plasma forming its outer layer eventually follow a path outside the sub-box.
Additionally, the pressure-strain interaction's rise and decay happen slightly before the electrical work rate's decrease and increase, respectively.

On Figure \ref{fig:local_nrj_transfer}(e) are the energy transfers rates temporal evolution around an X-point where two magnetic islands eventually coalesce, see Figures \ref{fig:NL_evolution_whole-box}, \ref{fig:NL_evolution_sub-box} and \ref{fig:rec_rate}.
At first, till $t = 580 \, \Omega_{ci}^{-1}$, there is primary reconnection ongoing around the X-point.
As such, the energy conversion rates are both positive, similarly to the ones shown in panel (d) although with more perturbed profiles. 
For $t > 585 \, \Omega_{ci}^{-1}$, as the islands are being pushed against one another, there is more heating and $\bm{j}_i \cdot \bm{E} < 0$ as there is a conversion back towards magnetic energy.
During this stage, the energy conversion rates profiles are more resemblant to those found for the center of a magnetic island in panel (c).
After some time ($t \simeq 625$), $\bm{j}_i \cdot \bm{E}$ becomes less negative because of the ongoing reconnection between the magnetic islands outer layers.

\begin{figure*}[t!] 
    \resizebox{\hsize}{!}{\includegraphics{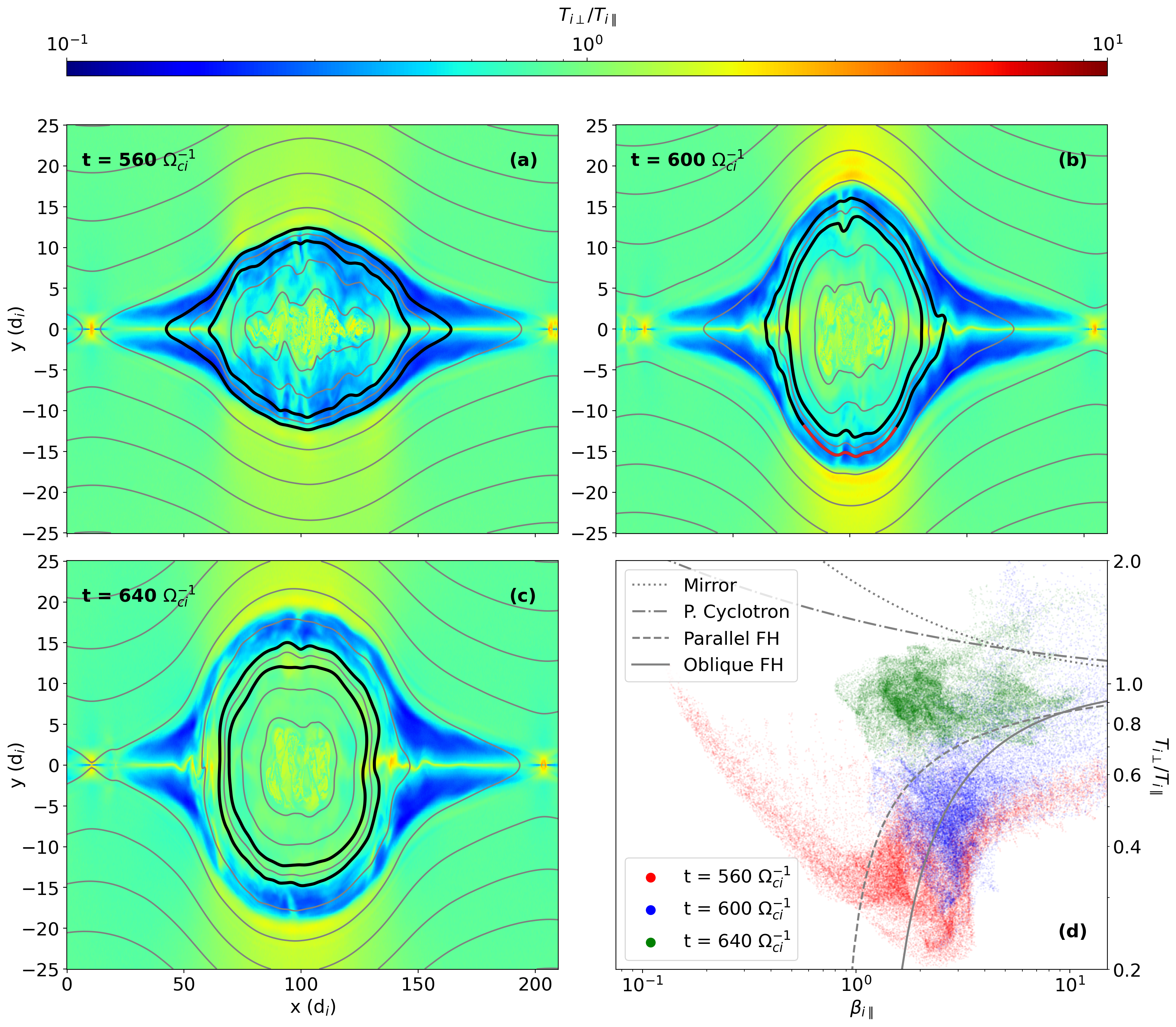}}
    \caption{\label{fig:FT_anisotropy_evolution}
        Evolution of the ion temperature anisotropy around a magnetic island.
        Panels (a), (b), and (c) show the spatial distribution of $T_{i \perp} / T_{i \parallel}$ at three different times: $t = 560 \, \Omega_{ci}^{-1}$, $t = 600 \, \Omega_{ci}^{-1}$ and $t = 640 \, \Omega_{ci}^{-1}$.
        The thicker plain black lines delimit the flux tube defined $A_z \in [5, 7] \, B_0 d_i$.
        Panel (c) gives the flux tube's plasma distribution, in the $\beta_{i \parallel} \text{-} (T_{i \perp} / T_{i \parallel})$ plane, for the three times shown in panels (a-c).
        Every dot corresponds to one simulation cell, and each color is associated to a different time.
        The grey curves correspond to marginal stability conditions for different ion kinetic instabilities.
    }
\end{figure*}  

\section{Temperature Anisotropy and Ion Firehose Instabilities} \label{sec:firehose_island}

We are now interested in mechanisms related to this energy conversion.
More precisely, this section focuses on the ion temperature anisotropy evolution, and firehose instabilities development during the non-linear stage of the tearing instability.

\subsection{Plasma Distribution and Temperature Anisotropy Evolution in a Magnetic Island}

Figure \ref{fig:FT_anisotropy_evolution}(a-c) display in color the ion temperature anisotropy $T_{i \perp} / T_{i \parallel}$ around the leftmost magnetic island in the simulation domain (located between X-points 1 and 2, see Figure \ref{fig:NL_evolution_whole-box}), for three different times $t = 560, 600$ and $640 \, \Omega_{ci}^{-1}$.
The times displayed are different from those in Figure \ref{fig:NL_evolution_whole-box} and Figure \ref{fig:NL_evolution_sub-box}, as they are chosen here to best highlight the temperature anisotropy evolution during the magnetic island's growth.
We note that, due to the chosen aspect ratio, the magnetic island in these panels appears less elongated along $x$ than what it is.
The reconnection outflows, and magnetic island outer layer, have a significant temperature anisotropy $T_{i \parallel} > T_{i \perp}$.
This anisotropy gets regulated ($T_{i \parallel} \sim T_{i \perp}$) from the inner to the outer layers of the plasmoid as it evolves.
Additionally, the reconnection jets temperature is more isotropic, and even exhibit $T_{i \perp} > T_{i \parallel}$, near their center, which is consistent with Speiser-like ion orbits in this region \citep{Speiser_1965_Orbits, Hietala_2015_reconnection_anisotropy_Speiser}.

\begin{figure}[t!] 
    \resizebox{\hsize}{!}{\includegraphics{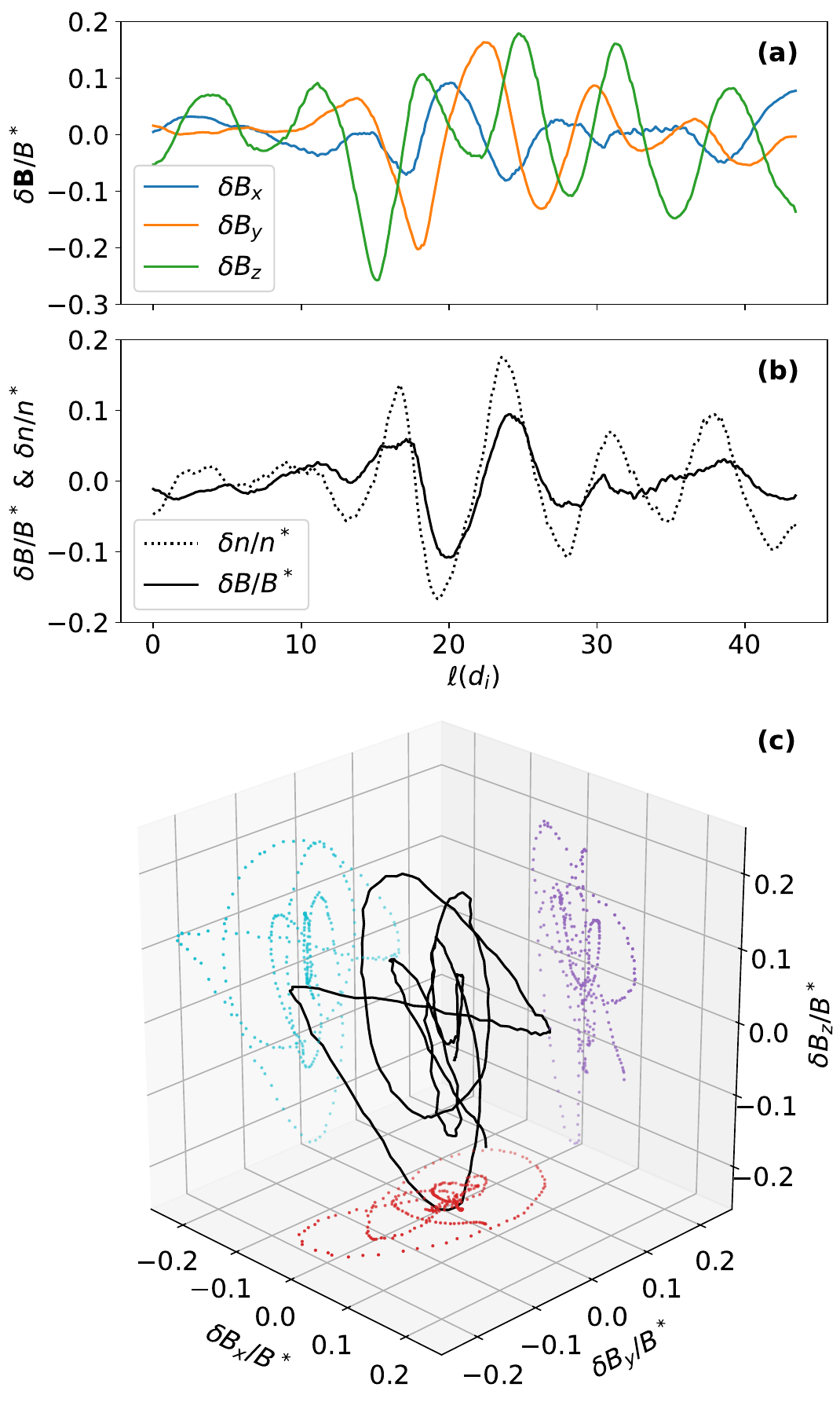}}
    \caption{\label{fig:FL_fluctuations}
        Magnetic field and density fluctuations at $t = 600 \, \Omega_{ci}^{-1}$, for a path between $x = 80 \, d_i$ and $x = 120 \, d_i$ and at $y < 0$ along the plasmoid field line defined by $A_z = 7 \, B_0 d_i$, as shown in red on Figure \ref{fig:FT_anisotropy_evolution}(b).
        Panel (a) shows fluctuations of the magnetic field components $\delta B_{x,y,z}$.
        Panel (b) gives fluctuations of the magnetic field's magnitude and plasma density, $\delta B$ and $\delta n$, respectively.
        Panel (c) displays the hodograms of $\delta B_{x,y,z}$, for all components in black and projections in colors.
    }
\end{figure}

In order to get more quantitative insights, we consider the plasma evolution within a single flux tube of the magnetic island.
The limits of the considered flux tube are indicated by the thick black lines in Figure \ref{fig:FT_anisotropy_evolution}(a-c).
These are defined by taking values of the vector potential $A_z \in [5, 7] \, B_0 \, d_i$, once the corresponding field lines have reconnected.
At $t =  515 \, \Omega_{ci}^{-1}$, the outer magnetic field line ($A_z = 7 \, B_0 \, d_i$) has just reconnected and lies in the direct vicinity of the X-points ($5 \, d_i$ for the left one and $15 \, d_i$ for the right one), so we consider times $t > 515 \, \Omega_{ci}^{-1}$.
The red part of the field line in panel (c) corresponds to the path along which we sample the fluctuations shown in Figure \ref{fig:FL_fluctuations}.

Figure \ref{fig:FT_anisotropy_evolution}(d) shows the temporal evolution, in the $\beta_{i \parallel} \, \text{-} \, T_{i \perp} / T_{i \parallel}$ plane, of the plasma distribution within the flux tube defined above.
The plasma distribution is displayed for three times $t = 560, \, 600$ and $640 \, \Omega_{ci}^{-1}$, corresponding to those of Figure \ref{fig:FT_anisotropy_evolution}(a-c), and distinguished by the different colored dots in red, blue and green, respectively.
We also show the linear marginal stability conditions for proton kinetic instabilities, at a growth rate $\gamma = 10^{-3} \, \Omega_{ci}$ in a homogeneous plasma with bi-Maxwellian proton velocity distribution functions \citep{Hellinger_2006_anisotropy_thresholds_SW}, by the different grey curves.
In the case of $T_{i \perp} > T_{i \parallel}$, these conditions for the proton cyclotron and mirror instabilities are indicated by the dashed-dotted and dotted grey curves.
For $T_{i \parallel} > T_{i \perp}$, the marginal stability conditions of the proton parallel and oblique firehose instabilities are given by the dashed and plain grey curves.
Although the reconnected plasma constituting the flux tube is not homogeneous with bi-Maxwellian proton velocity distribution functions, the marginal stability conditions still offer a reasonable estimation of when the plasma is expected to become unstable.
See Appendix \ref{sec-appendix:firehose_homogeneous} for results from a simulation of the firehose instabilities triggered in a homogeneous plasma.

At time $t = 560 \, \Omega_{ci}^{-1}$ (red dots in Figure \ref{fig:FT_anisotropy_evolution}(d)), the ion temperatures within the flux tube present important anisotropies $T_{i \parallel} > T_{i \perp}$.
Most of the plasma distribution exceeds the marginal stability conditions for the parallel ion kinetic firehose instability, with an important fraction of it also exceeding the oblique firehose stability condition.
The rest of the distribution is stable with regard to both instabilities, and presents an anticorrelation between $T_{i \perp} / T_{i \parallel}$ and $\beta_{i \parallel}$, which can be expected from a process dominantly increasing $T_{i \parallel}$ in these regions.

Later, for $t = 600 \, \Omega_{ci}^{-1}$ (blue dots), the large majority of the plasma distribution is exceeding either one, or both, ion kinetic firehose marginal stability conditions.
The previously stable part thus seems to have been driven to the unstable regions due to increase of $T_{i \parallel}$.
However, on average, the plasma has a more isotropic ion temperature and is distributed closer to the stability conditions.
The rest of the flux tube's plasma is sparsely distributed in the stable region and with $\beta_{i \parallel} > 1$, while very few points (at $\beta_{i \parallel} > 4$) exceed the ion cyclotron and mirror instability conditions.

Finally, at $t = 640 \, \Omega_{ci}^{-1}$ (in green) the ion temperature anisotropy has further reduced, such that most of the plasma distribution lies in the stable region.
Only a minority of points, located in the relatively high beta region $\beta_{i \parallel} \gtrsim 4$, exceed either the marginal stability conditions for the ion cyclotron and mirror, or kinetic firehose instabilities.
This indicates that a process, such as kinetic ion firehose instabilities or an analogous phenomenon, has eventually regulated (constrained) the anisotropy within the flux tube during the non-linear evolution of the tearing instability.

We now consider spatial fluctuations, following a path within $x \in [80, 120]$~$d_i$, and for $y < 0$, along the magnetic field line defined by $A_z = 7 B_0 d_i$, at $t = 600 \, \Omega_{ci}^{-1}$.
This path is indicated by the red curve on Figure \ref{fig:FT_anisotropy_evolution}(b).
We define local fluctuations between 1 and 10 $d_i$ of different quantities $a$ as $\delta a (\ell) = \langle a (\ell)  - \langle a (\ell) \rangle_{10 di} \rangle_{di}$, with $\ell$ the curvilinear abscissa along the field line, and brackets $\langle ... \rangle$ denoting local spatial averages over $\ell$.
Fluctuations of the different magnetic field vector components $\delta B_{x,y,z}$ are given on Figure \ref{fig:FL_fluctuations}(a), while fluctuations of the magnetic field magnitude $\delta B$ and of the plasma density $\delta n$ are shown on Figure \ref{fig:FL_fluctuations}(b).
Hodograms of $\delta B_{x,y,z}$ are displayed on Figure \ref{fig:FL_fluctuations}(c), as function of the three components in black, and projected on the different planes $\delta B_i$~-~$\delta B_j$ in colors.
The different quantities are normalized by either $B^*$ or $n^*$, the averages of $B$ and $n$ along the path.

On Figure \ref{fig:FL_fluctuations}(a), we observe that the dominant fluctuations are in $B_y$ and in $B_z$, so mostly transverse to the background magnetic field since $|B_x| / B^* \sim 0.95$, which is compatible with the ion kinetic firehose instabilities.
Fluctuations in $B_x$ are relatively small, with largest amplitude $| \delta B_x/B^*| \sim 0.1$ for $\ell \sim 20 \, \text{-} \, 25 \, d_i$.
Fluctuations $\delta B$ (panel (b)) are mostly anti-correlated with $\delta B_x$ (due to $B_x < 0$), and have similar amplitudes, with a maximum $\delta B /B^* \sim 0.1$ also for $\ell \sim 20 \, \text{-} \, 25 \, d_i$.
The density fluctuations $\delta n$, shown on panel (b), moreover exhibits larger normalized fluctuations ($|\delta n|/n^* \sim $~0.1~-~0.2), correlated to $\delta B$ (as in fast magnetosonic modes) for $\ell \in [12, 28] \, d_i$, which we attribute to the plasmoid compression.
Thus, the overall process is more complex than the ideal firehose instabilities from which are expected fluctuations transverse to the background field (so only $\delta B_y$ and $\delta B_z$) with a constant $|\mathbf{B}|$.
This is also highlighted by the non-trivial polarization displayed in the hodograms of Figure \ref{fig:FL_fluctuations}(c).
There is probably a superposition between a firehose-like process in a non-homogeneous plasma, and other fluctuations generated by the island compression.
A more precise characterization of the fluctuations would require further investigations, which are outside the scope of the present study.

\subsection{Average Plasma Evolution within a Magnetic Island Flux Tube} \label{sec:plasma_evol_FT}

\begin{figure}[t!] 
    \resizebox{\hsize}{!}{\includegraphics{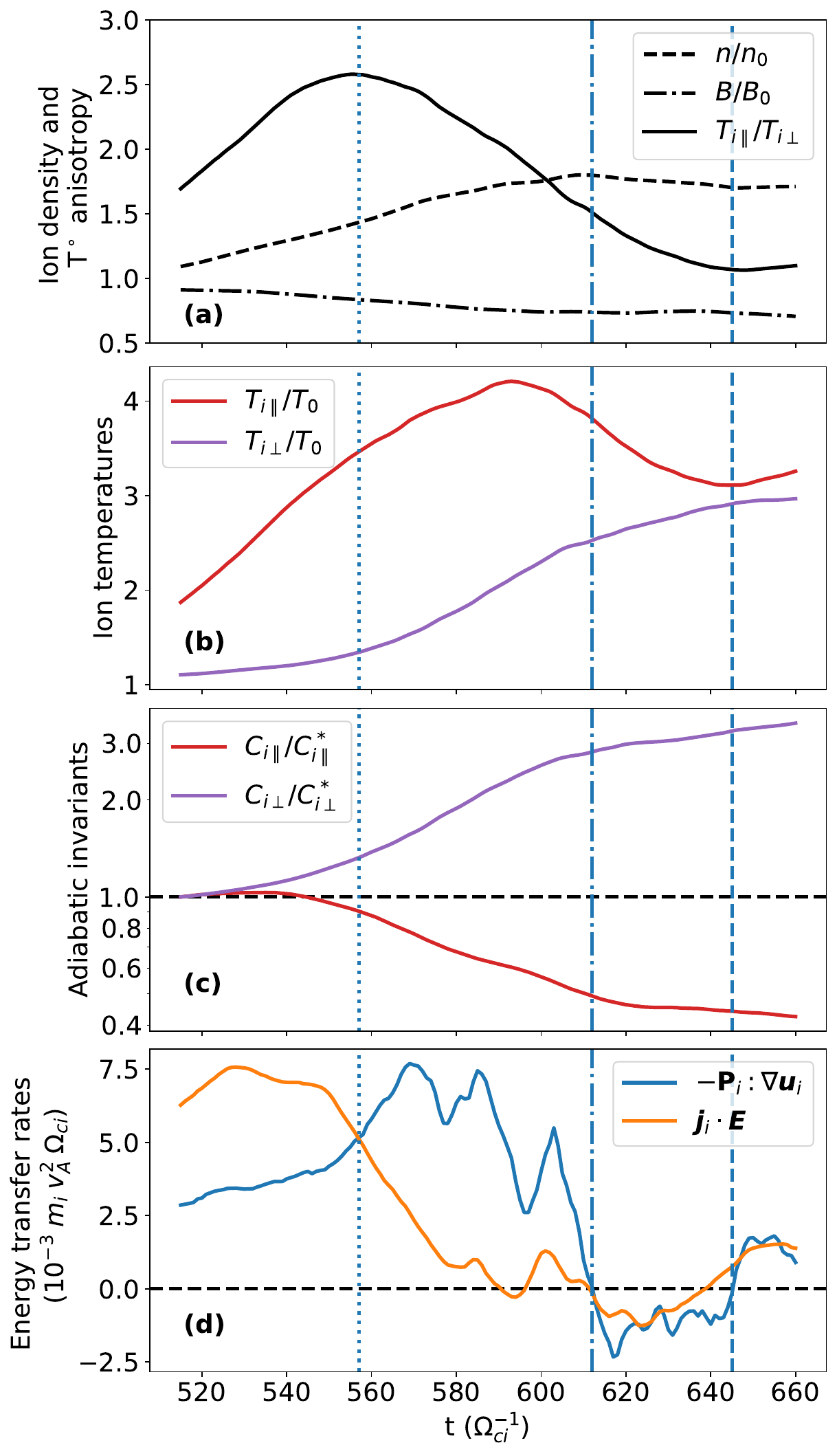}}
    \caption{\label{fig:FT_nrj_conversion_etc}
        Temporal evolution of several physical parameters spatially averaged over a magnetic island's flux tube.
        Panel (a) shows the normalized plasma density $n/n_0$, ion temperature anisotropy $T_{i \parallel} / T_{i \perp}$, and magnetic field magnitude $B/B_0$.
        Panel (b) gives the ion temperatures $T_{i \parallel}$ (red curve) and $T_{i \perp}$ (purple curve), while panel (c) provides the associated adiabatic invariants $C_{i \parallel}$ and $C_{i \perp}$ (see Equations (\ref{eq:adiab_invariants})), with matching colors.
        In panel (b), the temperatures $T_{i \parallel, \perp}$ are normalized by their initial isotropic value $T_0 = T_{i} (t=0)$, and in panel (c) the invariants $C_{i \parallel, \perp}$ are normalized to unity at the beginning of the interval by defining $C_{i \parallel, \perp}^* = C_{i \parallel, \perp} (t = 515 \, \Omega_{ci}^{-1})$.
        Panel (d) shows the ion electrical work rate $\bm{j}_i \cdot \bm{E}$ (orange curve) and pressure strain interaction $- \mathbf{P}_i : \bm{\nabla u}_i$ (blue curve).
        Vertical blue lines separate different physical regimes during the flux-tube evolution, see Section \ref{sec:plasma_evol_FT} for more details.
    }
\end{figure}  

In Figure \ref{fig:FT_nrj_conversion_etc} we give the temporal evolution of different physical parameters averaged within the flux tube delimited by the black plain lines in Figure \ref{fig:FT_anisotropy_evolution}(a-c).
The quantities $C_{i \parallel, \perp}$ in Figure \ref{fig:FT_nrj_conversion_etc}(c) correspond to the CGL \citep{Chew-Goldberger-Low_1956_adiabatic_invariants} ion adiabatic invariants, which are defined as
\begin{align} \label{eq:adiab_invariants}
    C_{i \parallel} = T_{i \parallel} \left( \frac{B}{n} \right)^2, && \text{and} && C_{i \perp} = \frac{T_{i \perp}}{B},
\end{align}
and are conserved ($D/Dt \, (C_{i \parallel, \perp}) = 0$) in the absence of heat flux and kinetic effects \citep{Chust_2006_fluid_eqs_closure}.
In the case of the ion firehose instabilities, we expect a non-conservation of these invariants, with an increase of $C_{i \perp}$, and a decrease of $C_{i \parallel}$, see Appendix \ref{sec-appendix:firehose_homogeneous}
The invariants in Figure \ref{fig:FT_nrj_conversion_etc}(c) have moreover been normalized by $C_{i \parallel, \perp}^* = C_{i \parallel, \perp} (t = 515 \, \Omega_{ci}^{-1})$, so that they are equal to unity at the beginning of the interval.

We note that there is a decrease of about 20-25 \% in the magnetic field magnitude, Figure \ref{fig:FT_nrj_conversion_etc}(a) during the entire flux tube evolution.
This overall participates to increasing $T_{i \parallel}$ and lessen the increase of $T_{i \perp}$, Figure \ref{fig:FT_nrj_conversion_etc}(b).
We describe below the evolution of the flux tube averaged plasma parameters for time intervals exhibiting different dynamics, and separated by the vertical lines in Figure \ref{fig:FT_nrj_conversion_etc}.

We first consider the interval going from $t = 515 \, \Omega_{ci}^{-1}$ to $t = 557 \, \Omega_{ci}^{-1}$ (vertical dotted line in Figure \ref{fig:FT_nrj_conversion_etc}).
During this stage, there is an overall compression of the plasma as can be seen from the increase of the average density $n$ in panel (a).
This compression is quasi-adiabatic as indicated by the $C_{i \parallel, \perp}$ behavior in panel (c).
Indeed, the parallel invariant $C_{i \parallel}$ is nearly constant, with only negligible variations, so there is an important increase of $T_{i \parallel}$ (panel (b)) due to the plasma compression.
There is also a slight increase of $C_{i \perp}$, coincident with a small increase of $T_{i \perp}$.
As a result, there is a significant rise of the temperature anisotropy $T_{i \parallel} / T_{i \perp}$ during this stage, see the black plain curve in panel (a).
Both the ion electrical work rate $\bm{j}_i \cdot \bm{E}$ and pressure-strain interaction term $- \mathbf{P}_i : \bm{\nabla u}_i$ are positive (see Figure \ref{fig:FT_nrj_conversion_etc}(d)), with $\bm{j}_i \cdot \bm{E} > - \mathbf{P}_i : \bm{\nabla u}_i$, implying an overall conversion of magnetic energy towards both ion bulk acceleration and heating.
At the end of the interval, $\bm{j}_i \cdot \bm{E} = - \mathbf{P}_i : \bm{\nabla u}_i$, so on average, all the magnetic energy being converted to the ions goes towards their internal energy (heating).
We note that around the magnetic island center, $\bm{j}_i \cdot \bm{E}$ was on average negative at all times (see Figure \ref{fig:local_nrj_transfer}(c)), whereas it is on average positive here.
This is due to the contribution of the high $\bm{j}_i \cdot \bm{E}$ positive values on the two lateral sides of the island near $y=0$, see Figure \ref{fig:local_nrj_transfer}(a), which are included within the flux tube averages shown in Figure \ref{fig:FT_nrj_conversion_etc} but outside the sub-box considered in Figure \ref{fig:local_nrj_transfer}(c).
From a kinetic point of view, this energy conversion and increase in $T_{i \parallel}$ is consistent with Fermi acceleration of the ions within the contracting island \citep{Drake_2010_proton_Fermi_acc_firehose}.

The second interval we consider starts at $t = 557 \, \Omega_{ci}^{-1}$ and ends at $t = 612 \, \Omega_{ci}^{-1}$, between the dotted and dashed-dotted vertical blue lines in Figure \ref{fig:FT_nrj_conversion_etc}, which includes the time considered on Figure \ref{fig:FL_fluctuations} ($t = 600 \, \Omega_{ci}^{-1}$).
The compression continues but it is no longer quasi-adiabatic.
There is a continuous decrease of $C_{i \parallel}$, implying a non-adiabatic source of parallel cooling which constraints the rise of $T_{i \parallel}$, and eventually makes it decrease around $t = 595 \, \Omega_{ci}^{-1}$.
We also observe an important similar increase in $C_{i \perp}$ and in $T_{i \perp}$, meaning that there is a significant non-adiabatic proton heating in the direction perpendicular to local magnetic field.
The ion temperature then progressively gets re-isotropized, as indicated by the decrease of $T_{i \parallel} / T_{i \perp}$ (panel (a)).
These signatures are consistent with the ion kinetic firehose instabilities, see Appendix \ref{sec-appendix:firehose_homogeneous}.
Regarding the energy conversion rates, there is, at the beginning of the interval, a sudden drop of $\bm{j}_i \cdot \bm{E}$ (panel (d)), which then remains very weak from $t \simeq 580 \, \Omega_{ci}^{-1}$ to $t = 612 \, \Omega_{ci}^{-1}$.
This implies that the magnetic energy conversion toward the ions is reduced until becoming nearly negligible.
Meanwhile, the ion pressure-strain interaction rises before decreasing, but remains significant through the whole interval, so there is continuous ion heating.
This heating comes at the expense of ion bulk kinetic energy because $\bm{j}_i \cdot \bm{E} < - \mathbf{P}_i : \bm{\nabla u}_i$. 

The third considered interval, between the dashed-dotted and dashed vertical lines, goes from $t = 612 \, \Omega_{ci}^{-1}$ to $t = 645 \, \Omega_{ci}^{-1}$.
The plasma compression stops during this stage, and $n/n_0$ even slightly decreases, see panel (a).
This evolution is still non-adiabatic, although to a lesser degree than in the previous stage.
In particular, both $C_{i \parallel}$ and $T_{i \parallel}$ continue to decrease while $C_{i \perp}$ and $T_{i \perp}$ continue to increase, see panels (b) and (c).
Hence, the average ion temperature anisotropy gets more and more constrained, until it reaches $T_{i \parallel} \simeq T_{i \perp}$ at the end of the interval.
Both energy conversion terms become slightly negative, implying that on average, ions cool down and transfer some of their energy (internal and/or kinetic) back towards the magnetic field.

After the third stage ($t > 645 \, \Omega_{ci}^{-1}$), the changes in plasma parameters become less abrupt.
These are associated with the long term evolution of the magnetic island, after the primary jet collision and associated dynamics, which goes out of the scope of the present study.

\section{\label{sec:conclusion} Discussion and Conclusion}

\subsection{Summary and discussion}


This study focuses on a 2.5 D hybrid simulation of the tearing instability using M\"obius periodic boundary conditions.
There is first a linear stage of the instability with an exponential growth of the lower wavenumbers and with several X-points along the current sheet.
The system then evolves towards a non-linear stage.
This appears to be associated with the onset of fast collisionless reconnection at the principal X-points, where a normalized reconnection rate of order $E_z^X / (v_{A,0} B_0) \sim 0.1$ is reached.

Although the non-linear stage of the tearing instability is often associated with fractal reconnection \citep[e.g.][]{Shibata_2001_fractal_reco, Daughton_2009_tearing_transition_collisional-kinetic, Ji_2011_phase_diagram_reconnection, Tenerani_2015_ideal_tearing_recursive, Landi_2015_resistive_MHD_ideal_recursive_reconnection, Papini_2019_sec_tearing_Hall-MHD}, it is not observed in the present simulation.
Since the current sheet thickness is of the order of $d_i$, this can be due to the Hall effect contribution \citep{Shi_2019_recursive_tearing_Hall}.
We propose that it may also be linked to the ion temperature anisotropy produced by the reconnection due to two reasons.
Firstly, an anisotropy with $T_{i \parallel} > T_{i \perp}$ tends to reduce the growth of the tearing instability \citep{Chen_1984_anisotropy_linear_tearing, Matteini_2013_tearing_anisotropic}.
Secondly, this anisotropy also reduces the effective magnetic tension, thus constraining the magnetic islands' contraction, with effectiveness increasing for higher plasma beta \citep{Schoeffler_2011_firehose_tearing_islands_growth}.
This then leads to less elongated secondary current sheets, which are therefore more stable with regard to the tearing instability.
This is further supported by the PIC-simulation study of \citet{Drake_2006_electron_acc_contracting_island}, where the authors show that the inclusion of a guide field (thus lowering the plasma beta) facilitates the production of secondary islands.
Kinetic effects, whether manifesting by a Hall electric field or anisotropic temperatures, therefore have to be considered for the development, or not, of a fractal hierarchy within a tearing unstable current sheet.
There might also be an effect of the finite spatial resolution.
In any case, more investigations are required to draw better conclusions. 

We find that the energy conversion is predominantly occurring during the tearing instability non-linear stage.
Globally, the magnetic energy continuously decreases, while the internal energy always increases, and the kinetic bulk speed energy increase before saturating due to the ouflows colliding.
Regarding the conversion rates, the transfer of magnetic energy increases after the start of the non-linear tearing instability, and decreases after some point as the reconnection slows down.
The heating follows a similar trend, but eventually increases again, which we attribute to the coalescence between two magnetic islands.

We also estimated the local energy conversion evolution at different places within the current sheet.
Around the X-points, there is nearly the same amount of magnetic energy going towards heating and bulk plasma outflow.
We find that there is overall more heating occurring within the magnetic islands than around the X-points, which agrees with results from MHD simulations conducted by \citet{Birn_2005_energy_conversion_reconnection}.
Regarding reconnection in a more general context, this implies that heating is not solely due to conversion in the vicinity of the X-point, but also ultimately linked to the jetting interaction with the rest of the plasma (e.g. collision with another structure).
Previous studies have shown that the contraction of magnetic islands' \citep{Drake_2006_secondary_islands, Schoeffler_2011_firehose_tearing_islands_growth}, and more generally compressibility in plasma turbulence \citep{Adhikari_2025_Helmholtz_decomposition_PS}, become more effective when lowering the plasma beta.
Therefore, the initial plasma beta might also have a significant influence on energy conversion during the non-linear tearing instability; further investigations would be required in order to evaluate this effect properly.
We moreover show that, after some point, the coalescence between magnetic island also participates to plasma heating.

The reconnection outflows exhibit an ion temperature anisotropy $T_{i \parallel} > T_{i \perp}$, which, after some time, gets regulated (constrained) within the magnetic islands.
We isolated a plasmoid flux tube and studied the overall plasma evolution within it, and found that it is consistent with a competition between two effects during the island contraction.
At first, the flux tube compression leads to an increase of $T_{i \parallel}$, hence driving and sustaining the ion temperature anisotropy.
Then, a non-adiabatic process, consistent with the firehose instabilities, eventually dominates over the compression, making $T_{i \parallel}$ decrease and $T_{i \perp}$ increase, thus reducing the anisotropy.
For an initially homogeneous plasma with bi-Maxwellian ion velocity distribution functions, the ion kinetic firehose instabilities operate in two major stages.
The first corresponding to the growth of transverse fluctuations, and the second to their damping, both are non-adiabatic and contribute to the temperature anisotropy regulation, see Appendix \ref{sec-appendix:firehose_homogeneous}.
In the present tearing simulation, this process is further complexified by the long-lasting driving of temperature anisotropy due to the plasmoid contraction, and the fact that the reconnected plasma is not homogeneous with bi-Maxwellian ion velocity distribution functions. 
Therefore, even though the plasma evolution is consistent with the firehose instabilities being the dominant processes responsible for the ion temperature re-isotropization, there might also be a contribution from other effects and non-linear phenomena.

A natural next step would be the incorporation of a guide field, and conduct a similar investigation for 3D simulations where fluctuations along the $z$ direction can develop, as these can significantly affect the tearing instability dynamics \citep[e.g.][]{Drake_2006_secondary_islands, Landi_2008_3D_tearing_MHD_simu, Markidis_2013_3D_tearing_PIC, Cerutti_2014_tearing_3D}.
A 3D simulation would further improve the resolution of kinetic processes, such as small-scale turbulence and firehose instabilities, whose coupling is more accurately described in a three-dimensional framework \citep{Hellinger_2019_3D_FH_HEB, Gary_2020_2D-3D_instabilities-turbulence}.
It would moreover be interesting to inspect and characterize the ion velocity distribution functions evolution, as well as the production of energetic particles. 

Full PIC simulation can also give to access to the kinetic electron dynamics, which are not available in the hybrid-PIC approximation.
These allow to include potentially important effects of the electrons on the reconnection evolution \citep{Shay_2025_review_simu_models_reconnection}, as well as studying the interplay between protons and electrons and how the energy is distributed between the two species \citep[e.g.][]{Yin-Drake-Swisdak_2025_electron-proton_heating_tearing}.
There are for example evidences that reconnection can also induce parallel electron temperature anisotropy, and trigger the associated kinetic firehose instability \citep{Le_2019_electron_firehose_simu_3D_PIC, Cozzani_2023_elec_firehose_reconnection_MMS}.
This could be relevant for the ions' dynamics since the electron temperature anisotropy affects both branches of the proton firehose instability \citep{Kennel-Scarf_1968_proton_firehose, Hellinger-Matsumoto_2000_oblique_FH, Maneva_2016_electron_anisotropy_oblique_proton_FH}.

%

\subsection{Conclusions}

Here are our principal conclusions regarding the tearing instability simulation presented in this study.
M\"obius boundary conditions allow to divide the simulation domain by two as compared to regular periodic boundaries, and thus double the computation efficiency for the same results.
Energy conversion dominantly occurs during the non-linear stage of the tearing instability, and is highly inhomogeneous.
In the X-points vicinity, nearly the same amount of magnetic energy is transferred towards plasma acceleration than into ion heating.
Most of the heating however occurs within the magnetic islands, which dynamics are therefore crucial for energy conversion.
These dynamics are intrinsically linked to the kinetic firehose instabilities, regulating the temperature anisotropy ($T_{i \parallel} > T_{i \perp}$) generated in the reconnected plasma, and participating in non-trivial ways for the energy conversion through generation and damping of plasma waves/fluctuations.
We further propose that the absence of fractalization in our simulation is also due to the $T_{i \parallel} > T_{i \perp}$ temperature anisotropy, and highlights the importance of kinetic effects on a global scale.

\begin{acknowledgments}

E. B. wishes to acknowledge valuable comments from M. Velli and B. Lavraud.
The authors acknowledge support from the collaboration project  ``Energetics of tearing-driven reconnection in weakly collisional plasmas'' between the CNRS under the call ``International Emerging Actions (IEA)" (2024) and  the Czech Academy of Sciences (CAS) under the call ``Mobility Plus Project'' (CNRS-25-06).
This work was granted access to the HPC resources of MesoPSL financed by the Region Ile de France and the project Equip@Meso (reference ANR-10-EQPX-29-01) of the programme Investissements d’Avenir supervised by the Agence Nationale pour la Recherche.

\end{acknowledgments}

\appendix

\section{Energy Conversion during the Ion Kinetic Firehose Instabilities in an Homogeneous Plasma} \label{sec-appendix:firehose_homogeneous}

\begin{figure*}[t!] 
    \resizebox{\hsize}{!}{\includegraphics{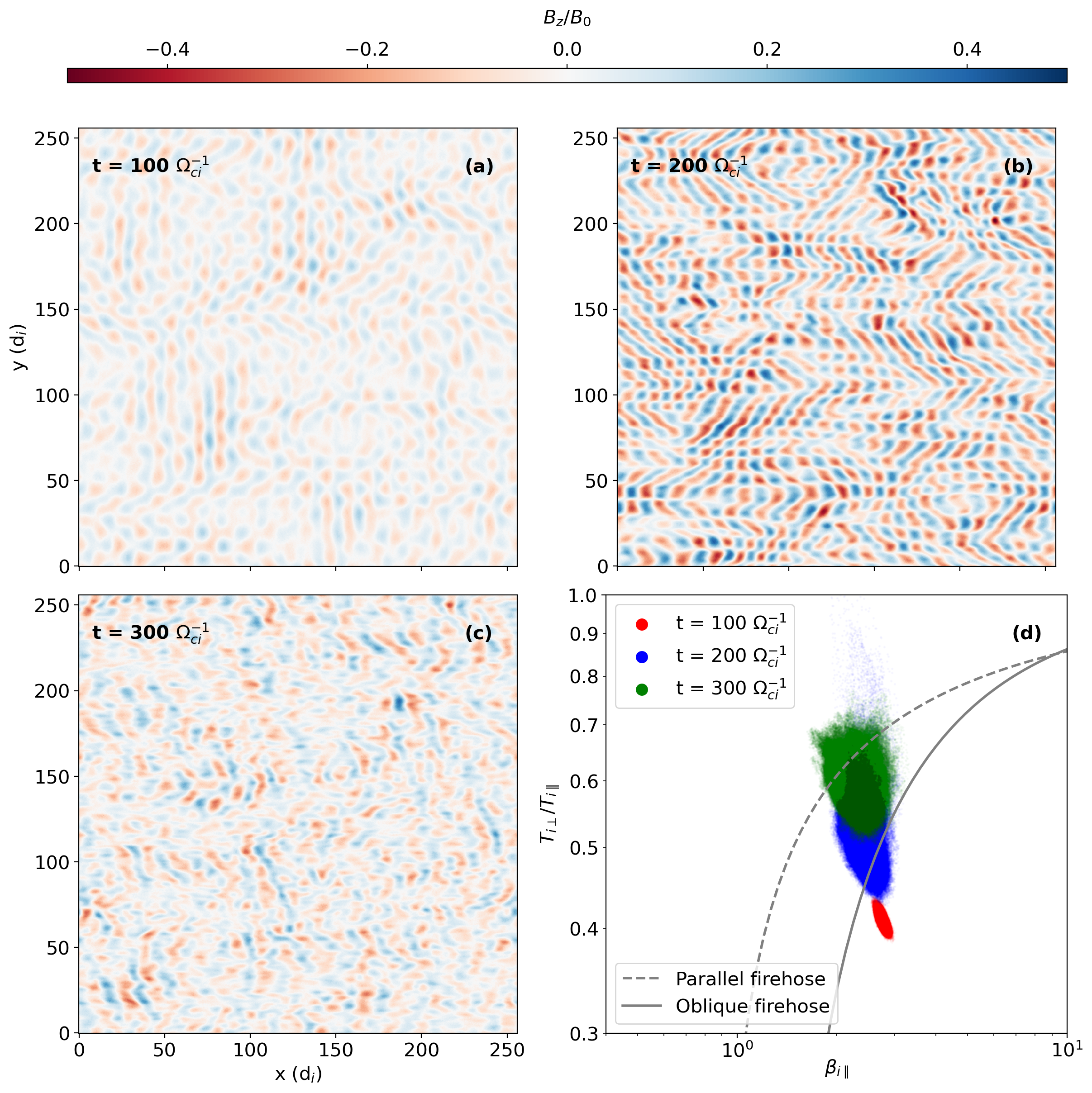}}
    \caption{\label{fig:FH_anisotropy_evolution}
        General evolution of magnetic field fluctuations and ion temperature anisotropy during the kinetic firehose instabilities triggered in an initially homogeneous plasma with bi-Maxwellian ion velocity distribution functions.
        Panels (a), (b) and (c) display the out-of-plane magnetic field component $B_z$ for three different times: $t = 100 \, \Omega_{ci}^{-1}$, $t =200 \, \Omega_{ci}^{-1}$ and $t = 300 \, \Omega_{ci}^{-1}$.
        Panel (c) shows the plasma distribution in the $\beta_{i \parallel} \, \text{-} \, T_{i \perp} / T_{i \parallel}$ plane, for the three times shown in panels (a-c), distinguished by different colored dots.
    }
\end{figure*}  

In order to get better insights of kinetic firehose instabilities effects on energy conversion, this section presents results of another 2.5 D hybrid simulation, starting in an homogeneous plasma with bi-Maxwellian ion velocity distribution functions.
The simulation is initialized with an ion temperature anisotropy $T_{i \perp} / T_{i \parallel} = 0.4$, a parallel ion beta $\beta_{i \parallel} = 2.8$, and a background magnetic field along the $x$ direction $\bm{B}_0 = B_0 \, \bm{e}_x$.
These are the same parameters as in the study of \citet{Hellinger_2001_parallel_oblique_FH_NL_competition}.
The simulation domain has a total size of $1024^2$ cells, with a grid resolution of $\Delta x = \Delta y = 1/4 \, d_i$, and 4096 particles per cell.
As for the tearing simulation, we do not include any initial perturbation so the instability starts to grow from the noise.
Below, we summarize the general behavior of the instability \citep[see also][]{Hellinger-Matsumoto_2000_oblique_FH, Hellinger_2001_parallel_oblique_FH_NL_competition} and quantify the evolution of some relevant parameters, including the energy conversion rates.

We give an overview of the instability evolution, and associated ion temperature isotropization, in Figure \ref{fig:FH_anisotropy_evolution}.
In Figure \ref{fig:FH_anisotropy_evolution}(a-c), the out-of-plane magnetic component $B_z$ is shown at three different times.
We can see fluctuations progressively appear ($t = 100 \, \Omega_{ci}^{-1}$, panel (a)), grow ($t = 200 \, \Omega_{ci}^{-1}$, panel (b)) and finally decay ($t = 300 \, \Omega_{ci}^{-1}$, panel (c)) during the simulation.
Meanwhile, the plasma distribution in the $\beta_{i \parallel} \, \text{-} \, T_{i \perp} / T_{i \parallel}$ plane (panel (d)) gradually evolves away from the unstable region, within the kinetic firehose marginal stability conditions (grey curves).
The distribution indeed goes from being in the unstable region for both instabilities at $t = 100 \, \Omega_{ci}^{-1}$ (red dots), to being in the stable region, or only unstable with regard to the parallel firehose instability at $t = 300 \, \Omega_{ci}^{-1}$ (green dots), and with a broad distribution of points.

\begin{figure}[t!] 
    \resizebox{\hsize}{!}{\includegraphics{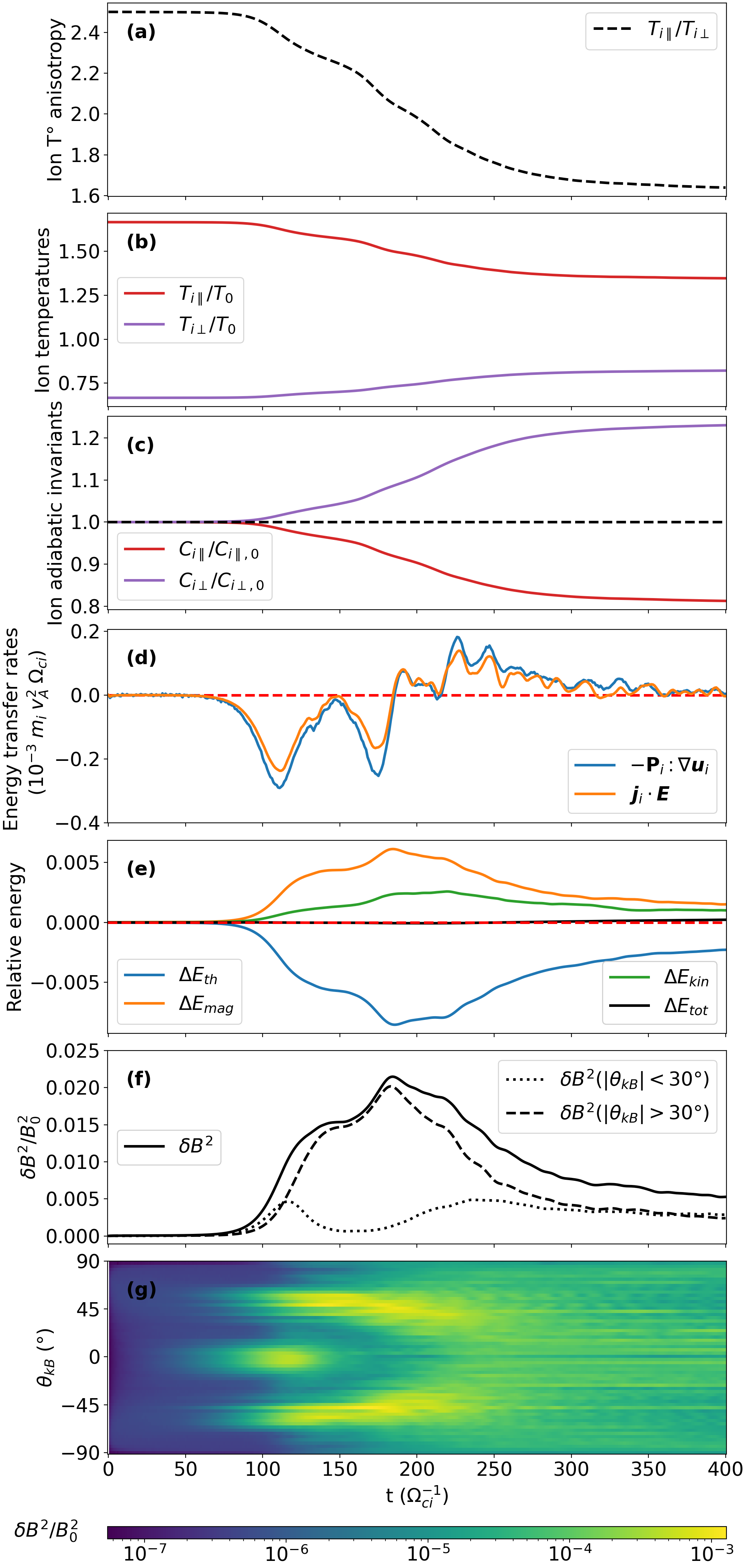}}
    \caption{\label{fig:FH_nrj_conversion_etc}
        Spatially averaged parameters evolution during a simulation of the ion kinetic firehose instabilities.
        Panels (a-d) are the same as in Figure \ref{fig:FH_nrj_conversion_etc}: ion temperature anisotropy $T_{i \perp} / T_{i \parallel}$ (a) and values $T_{i \parallel, \perp}$ (b), adiabatic invariants $C_{i \parallel, \perp}$ (c), and energy conversion rates $\bm{j}_i \cdot \bm{E}$ and $- \mathbf{P}_i : \bm{\nabla u}_i$ (d).
        Panel (e): relative energies evolution $\Delta E$, as in Figure \ref{fig:nrj_PS_etc}(a).
        Panel (f): magnetic fluctuations $\delta B^2 / B_0^2$, for $|\theta_{kB}| \gtrless 30^\circ$ (dashed and dotted curves) and total (plain curve).
        Panel (g): evolution of $\delta B^2 / B_0^2$ (colors) as function of $\theta_{kB}$, integrated over $3^\circ$ bins.
    }
\end{figure}  

Figure \ref{fig:FH_nrj_conversion_etc} shows the evolution of several parameters spatially averaged over the domain.
We can see on panel (a) the temperature anisotropy $T_{i \perp} / T_{i \parallel}$ decreasing during the simulation, more particularly from $t \simeq 100 \, \Omega_{ci}^{-1}$ to $t \simeq 300 \, \Omega_{ci}^{-1}$, as the changes are fairly negligible outside this time interval.
This re-isotropization is due to a decrease of $T_{i \parallel}$ and an increase of $T_{i \perp}$, both also predominantly within the time range $t \in [100, 300] \, \Omega_{ci}^{-1}$.
As expected, the adiabatic invariants $C_{i \parallel, \perp}$ (normalized here by $C_{i (\parallel, \perp), 0} = C_{i \parallel, \perp} (t=0)$) follow the same trend as the associated temperatures.
This regulation of temperature anisotropy follows two major stages, which we describe below.

Firstly, for $t < 190 \, \Omega_{ci}^{-1}$, both energy transfer rates $- \mathbf{P}_i : \bm{\nabla u}_i$ and $\bm{j}_i \cdot \bm{E}$ are negative (significantly only after $t = 100 \, \Omega_{ci}^{-1}$), with $- \mathbf{P}_i : \bm{\nabla u}_i \lesssim \bm{j}_i \cdot \bm{E}$.
There is thus an increase of magnetic energy $\Delta E_{mag}$ and, to a lesser degree, of kinetic energy $\Delta E_{kin}$, while the ion internal energy $\Delta E_{th}$ decreases, see Equations (\ref{eq:EM_nrj_conv_avg}-\ref{eq:thermal_nrj_conv_avg}).
Secondly, after $t = 190 \, \Omega_{ci}^{-1}$, both $- \mathbf{P}_i : \bm{\nabla u}_i$ and $\bm{j}_i \cdot \bm{E}$ become positive and their amplitudes remain close, mostly with $- \mathbf{P}_i : \bm{\nabla u}_i \gtrsim \bm{j}_i \cdot \bm{E}$.
The plasma is thus heated and $\Delta E_{th}$ increases while there is a decrease of $\Delta E_{mag}$ and $\Delta E_{kin}$ (here also less than $\Delta E_{mag}$).

We can moreover see on Figure \ref{fig:FH_nrj_conversion_etc}(f) that the total magnetic fluctuations (black plain line) amplitude increase until $t = 190 \, \Omega_{ci}^{-1}$ and then decrease after.
Our interpretation for the global behavior of the system is thus as follows.
There is a first stage corresponding to an emergence of fluctations during the growth of instability.
Then, during a second stage the fluctuations are damped and redistribute their magnetic and kinetic energies into ion internal energy.
Both stages are non-adiabatic and contribute to reducing the temperature anisotropy by reducing $T_{i \parallel}$ and increasing $T_{i \perp}$.

During the first stage ($t < 190 \, \Omega_{ci}^{-1}$), the energy conversion rates profiles (panel (d)) have moreover the same shape with two dips, peaking around $t = 110 \, \Omega_{ci}^{-1}$ and $t = 175 \, \Omega_{ci}^{-1}$, and are nearly zero around $t = 150 \, \Omega_{ci}^{-1}$.
On panel (g), we can see fluctuations appearing at parallel ($| \theta_{kB} | \in [0, 15]^\circ$) and oblique ($| \theta_{kB} | \in [35, 65]^\circ$) wavevectors around $t = 50 \, \Omega_{ci}^{-1}$.
Then, $\delta B^2 (| \theta_{kB} | < 30^\circ)$ and $\delta B^2 (| \theta_{kB} | > 30^\circ)$ grow until $t \sim 110 \, \Omega_{ci}^{-1}$, after which $\delta B^2 (| \theta_{kB} | < 30^\circ)$ starts to decrease while $\delta B^2 (| \theta_{kB} | > 30^\circ)$ continues to increase.
This corresponds to the first dip shape in the energy conversion rates profiles.
The increase (or decrease) of $\delta B^2 (| \theta_{kB} | < 30^\circ)$ and $\delta B^2 (| \theta_{kB} | > 30^\circ)$ are associated to excitation (or damping) of the parallel and oblique branches of the ion firehose instability.
There is then a stage, around $t = 150 \, \Omega_{ci}^{-1}$, where the total $\delta B^2 \sim cst$ because both $\delta B^2 (| \theta_{kB} | < 30^\circ)$ and $\delta B^2 (| \theta_{kB} | > 30^\circ)$ reach a plateau, and as such $\mathbf{P}_i : \bm{\nabla u}_i \simeq \bm{j}_i \cdot \bm{E} \simeq 0$.
Fluctuations at oblique wavevectors $\delta B^2 (| \theta_{kB} | > 30^\circ)$ then grow again and attain a maximum for $t < 190 \, \Omega_{ci}^{-1}$, while $\delta B^2 (| \theta_{kB} | < 30^\circ) \sim cst$.
This corresponds to the second energy conversion rates dip, and we can see on panel (g) that it is associated with oblique fluctuations drifting towards more parallel wavevectors, while gaining some energy in the process.

During the second stage, after $t < 190 \, \Omega_{ci}^{-1}$, both $\mathbf{P}_i : \bm{\nabla u}_i$ and $\bm{j}_i \cdot \bm{E}$ are almost always positive and slowly decreasing, while also presenting some oscillatory behavior.
Until $t \simeq 240 \, \Omega_{ci}^{-1}$, fluctuations $\delta B^2 (| \theta_{kB} | > 30^\circ)$ decrease while those at $\delta B^2 (| \theta_{kB} | < 30^\circ)$ increase.
This is because the more oblique fluctuations continue to be converted towards more parallel wavevectors, as can be seen on Figure \ref{fig:FH_nrj_conversion_etc}(g) where their spectrum drifts towards lower $| \theta_{kB} |$ values.
However, the fluctuations total $\delta B^2$ decreases, implying some damping during this process, which explains the energy conversion rates positive values.
After $t \simeq 240 \, \Omega_{ci}^{-1}$, both $\delta B^2 (| \theta_{kB} | > 30^\circ)$ and $\delta B^2 (| \theta_{kB} | < 30^\circ)$ decrease, meaning that the damping has become dominant over the waves conversion from oblique to parallel wavevectors.
Then, with time, the fluctuations decay less and less and the energy conversion rates decrease.
Thus, the damping, while still remaining the dominant process, becomes steadily less effective.
Finally, although it would probably have continued to decrease slowly, a significant part of $\delta B^2$ ($\sim 25 \%$ as compared to the maximum) still remains by the end of the simulation, as already remarked by \citet{Hellinger_2001_parallel_oblique_FH_NL_competition}.

Finally, although the simulations have same initial physical parameters, there are some differences with the study of \citet{Hellinger_2001_parallel_oblique_FH_NL_competition}.
Most notably, the relative fluctuations energy at oblique to parallel wavevectors is higher in our case than in \citet{Hellinger_2001_parallel_oblique_FH_NL_competition}.
We attribute this principally to the domain's dimension, $L_x \times L_y = 256 \times 256 \, d_i$ here as compared to $L_x \times L_y = 256 \times 128 \, d_i$ in \citet{Hellinger_2001_parallel_oblique_FH_NL_competition}, allowing the growth of more oblique wavevectors due to a larger extension in the $y$ direction.
Some finer differences might also be due to the different numbers of macroparticles per cell (256 versus 4096) and spatial resolution ($d_i$ versus $d_i / 4$).







\bibliography{references}{}
\bibliographystyle{aasjournalv7}

\end{document}